\documentclass[aps,prd,twocolumn,nofootinbib,superscriptaddress,letterpaper]{revtex4}
\usepackage[hypertex]{hyperref}
\usepackage{graphicx}
\def\lesssim{\mathrel{\mathpalette\vereq<}}

\begin{document}

\pagebreak	gestyle{plain}

\title{Folded Supersymmetry and the LEP Paradox}
\author{Gustavo Burdman}
\affiliation{Instituto de F\'{i}sica, Universidade de S\~{a}o Paulo, \\
R. do Mat\~{a}o 187, S\~{a}o Paulo, SP 05508-0900, Brazil} 
\author{Z. Chacko}
\author{Hock-Seng Goh}
\affiliation{Department of Physics, University of Arizona, Tucson, AZ 85721}
\author{Roni Harnik}
\affiliation{
SLAC, Stanford University, Menlo Park, CA 94025\\
Physics Department, Stanford University,
Stanford, CA 94305}


\begin{abstract}

We present a new class of models that stabilize the weak scale against radiative corrections up to scales 
of order 5 TeV without large corrections to precision electroweak observables. In these `folded 
supersymmetric' theories the one loop quadratic divergences of the Standard Model Higgs field are cancelled 
by opposite spin partners, but the gauge quantum numbers of these new particles are in general different 
from those of the conventional superpartners. This class of models is built around the correspondence that 
exists in the large $N$ limit between the correlation functions of supersymmetric theories and those of 
their non-supersymmetric orbifold daughters. By identifying the mechanism which underlies the cancellation 
of one loop quadratic divergences in these theories, we are able to construct simple extensions of the 
Standard Model which are radiatively stable at one loop. Ultraviolet completions of these theories can be 
obtained by imposing suitable boundary conditions on an appropriate supersymmetric higher dimensional 
theory compactified down to four dimensions. We construct a specific model based on these ideas which 
stabilizes the weak scale up to about 20 TeV and where the states which cancel the top loop are scalars not 
charged under Standard Model color. Its collider signatures are distinct from conventional supersymmetric 
theories and include characteristic events with hard leptons and missing energy.

\end{abstract}

 \maketitle


\section{Introduction}

Precision electroweak measurements performed at LEP over the past decade, while lending strong
support to the Standard Model (SM), have lead to an apparent paradox~{\cite{LEPparadox}}. These
experiments are completely consistent with
\begin{itemize}
\item
the existence of a light SM Higgs with mass less than about 200 GeV, and also
\item
a cutoff $\Lambda$ for non-renormalizable operators that contribute to the precision
electroweak observables greater than or of order 5 TeV.
\end{itemize}
The problem arises because quadratically divergent loop corrections from scales of order 5 TeV, particularly 
from diagrams involving the top quark, naturally generate a Higgs mass much larger than 200 GeV in the 
SM.  This is called the `LEP paradox'.

The LEP paradox seems to suggest the existence of new physics at or below a TeV that cancels
quadratically divergent contributions to the Higgs mass, but does not contribute significantly to 
precision electroweak observables. One interesting possibility is weak scale supersymmetry, where R-parity 
ensures that contributions to precision electroweak observables are small. Here the quadratically divergent 
contributions to the Higgs mass from the top quark are cancelled by new diagrams involving the scalar 
stops, shown in Figure~(\ref{fig-intro}).
 
Little Higgs theories~{\cite{Little1, Little2}} constitute another approach to the LEP paradox. Models of 
this type with a custodial SU(2)~{\cite{custodial}} and T-parity~{\cite{Tparity}} do not give large 
corrections to precision electroweak observables. Warped extra-dimensional realizations of the Higgs as a 
pseudo-Goldstone boson~{\cite{CNP} are closely related to little Higgs theories. Reviews of this class of 
models and more references may be found in {\cite{Reviews}. In little Higgs theories the top loop is 
cancelled by diagrams involving new fermions, the `top-partners', which are charged under color and whose 
couplings to the Higgs are related by symmetry to the top Yukawa coupling. These diagrams are also shown in 
Figure~(\ref{fig-intro}).

Recently twin Higgs theories~{\cite{twin}}{\cite{twinLR}}, (see also~{\cite{CHWtwin},{\cite{FPStwin}}), a 
new class of solutions to the LEP paradox, have been proposed. These models have the feature that the 
diagrams which cancel the top loop have exactly the same form as in little Higgs theories, but the 
top-partners are not necessarily charged under SM color. The reason is that in a twin Higgs theory the 
top-partners need be related to the SM top quarks only by a discrete symmetry and not by a global symmetry as 
in little Higgs theories, and so do not necessarily carry the same color charge. Clearly, what is crucial 
for the cancellation to go through is that the couplings of the top-partners to the Higgs be related by 
symmetry in a specific way to the top Yukawa coupling. In these diagrams color serves merely as a 
multiplicity factor, and therefore whether the top-partners are charged under SM color or not 
is irrelevant to the cancellation.
\begin{figure} \begin{center} 
\includegraphics[width=0.85\columnwidth]{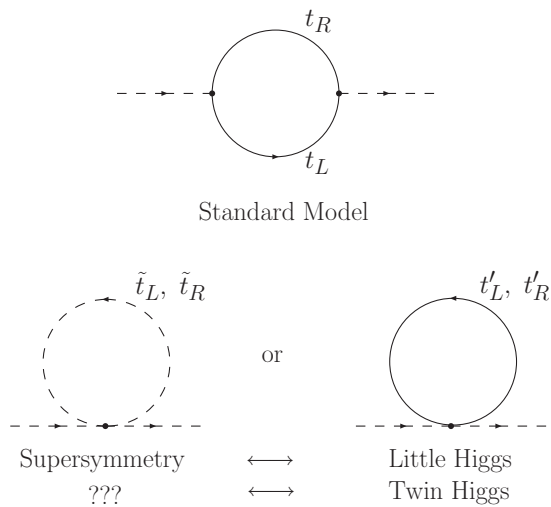} 
\end{center} \caption{
The diagram on top shows the contribution to the Higgs mass squared parameter in the SM from the top loop, 
while the lower two diagrams show how this contribution is cancelled in supersymmetric theories and in 
little Higgs theories. In twin Higgs models the cancellation takes place through a diagram of the same form 
as in the little Higgs case but the particles running in the loop need not be charged under color. In 
analogy with this, we seek a theory where the cancellation takes the same form as in the supersymmetric 
diagram but the states in the loop are not charged under color.}
\label{fig-intro} 
\end{figure}

At this point we turn our attention back to the supersymmetric case, where the cancellation of the top loop 
is realized by the scalar stops. Note that the fact that the stops are charged under SM color does not seem 
crucial for this cancellation, any more than in the little Higgs case. As before, color seems to serve 
merely as a multiplicity factor and what is necessary for the cancellation to go through is that the 
couplings of the scalars to the Higgs be related by symmetry in a specific way to the top Yukawa coupling. 
This observation begs the following question. Do there exist realistic theories where the quadratic 
divergence from the top loop is cancelled by a diagram of the same form as in the supersymmetric case, but 
where the scalars running in the loop are not charged under SM color?

The purpose of this paper is to answer this question firmly in the affirmative. We will construct a 
realistic model where the top loop is cancelled by scalars not charged under color. Moreover, in doing so 
we 
will go much further and outline the general construction of simple extensions of the SM where one loop 
quadratically divergent contributions to the Higgs mass from gauge and Yukawa interactions are cancelled by 
opposite spin partners whose gauge quantum numbers can in principle be very different from those of the 
conventional superpartners. We expect these results to enable the construction of entirely new classes of 
models that address the LEP paradox.

Our starting point is the observation that in the large $N$ limit a relation exists between the correlation 
functions of a class of supersymmetric theories and those of their non-supersymmetric orbifold daughters 
that holds to all orders in perturbation theory {\cite{fold1, fold2, fold3, fold4}}. The masses of scalars 
in the daughter theory are protected against quadratic divergences by the supersymmetry of the mother 
theory. In many cases the correspondence between the mother and daughter theories continues to hold to a 
good approximation even away from the large $N$ limit. By understanding the dynamics which underlies this 
cancellation, we can construct simple non-supersymmetric extensions of the SM where the Higgs mass is 
protected from large radiative corrections at one loop.{\footnote{For an earlier approach to stabilizing 
the weak scale also based on the large $N$ orbifold correspondence see \cite{Frampton}.}} These theories 
stabilize the weak scale against radiative corrections up to about 5 TeV, thereby addressing the LEP 
paradox.

In general, the low energy spectrum of such a `folded supersymmetric' theory is radically different from 
that of a conventional supersymmetric theory, and the familiar squarks and gauginos need not be present.  
While the diagrams that cancel the one loop quadratically divergent contributions to the Higgs mass have 
exactly the same form as in the corresponding supersymmetric theory, the gauge quantum numbers of the 
particles running in the loops, the `folded superpartners' (or `F-spartners' for short), need not be the 
same. This means that the characteristic collider signatures of folded supersymmetric theories tend to be 
distinct from those of more conventional supersymmetric models.

A folded supersymmetric theory does not in general possess any exact or approximate symmetry that 
guarantees that the form of the Lagrangian is radiatively stable. It is therefore particularly important to 
understand if ultraviolet completions of these theories exist. We show that supersymmetric ultraviolet 
completions where corrections to the Higgs mass from states at the cutoff are naturally small can be 
obtained by imposing suitable boundary conditions on an appropriate supersymmetric higher dimensional 
theory compactified down to four dimensions. We investigate in detail one specific model constructed along 
these lines. While in this theory the one loop radiative corrections to the Higgs mass from gauge loops are 
cancelled by gauginos, the corresponding radiative corrections from top loops are cancelled by particles 
not charged under SM color. In such a scenario the familiar supersymmetric collider signatures associated 
with the decays of squarks and gluinos that have been pair produced are absent. Instead, the signatures 
include events with hard leptons and missing energy that can potentially be identified at the LHC.

This paper is organized as follows. In the next section we explain the basics of orbifolding supersymmetric 
large $N$ theories to non-supersymmetric ones and give some simple examples establishing the absence of one 
loop quadratically divergent radiative corrections to scalar masses in the daughter theories. Based on 
these examples we then identify the underlying dynamics behind these cancellations, and explain how to 
extend these results to construct larger classes of theories where one loop quadratic divergences are also 
absent. In section~\ref{applications} we apply these methods to show how the quadratic divergences of the 
Higgs in the SM can be cancelled, and outline ultraviolet completions of these theories based on 
Scherk-Schwarz supersymmetry breaking on higher dimensional orbifolds. In section~\ref{realistic} we 
present a realistic ultraviolet complete model based on these ideas and briefly discuss its phenomenology.

\section{Cancellation of Divergences in Orbifolded Theories}
\label{orbifolds}

What is the procedure to orbifold a parent supersymmetric field theory? First, identify a discrete symmetry 
of the parent theory. In order to obtain a non-supersymmetric daughter theory this discrete symmetry should 
be an R symmetry. Now `orbifolding' simply consists of eliminating all fields of the parent theory that are 
not invariant under the discrete symmetry. The interactions of the daughter theory are inherited from the 
Lagrangian of the parent theory by keeping all terms which involve only the daughter fields. We will begin 
by demonstrating this procedure in two examples, one with gauge interactions and one with Yukawa 
interactions. Then, in subsection~\ref{bifold} we will identify the mechanism that guarantees the 
cancellation of divergences at one loop and list the `rules' for building models where such a cancellation 
is realized.

\subsection{Examples of Orbifold Theories}\label{example} 
To clarify this procedure we apply it to orbifold a supersymmetric U(2$N$) gauge theory with 2$N$ flavors 
down to a non-supersymmetric daughter theory with a U($N$) $\times$ U($N$) gauge symmetry. The SU(2$N$) and 
U(1) component gauge fields of U(2$N$) are assumed to have the same strength when their generators are 
normalized appropriately. The U(2$N$) theory is invariant under a discrete $Z_2$ symmetry which is an 
element of the gauge group and is generated by a matrix $\Gamma$ that has the form
\begin{equation}
\label{Gamma}
\pmatrix{+1 & & & & & \cr
          & \ddots & & & & \cr
        & & +1  & & & \cr
        & & &-1 & & \cr
        & & & &\ddots & \cr
        & & & & & -1}
\end{equation}
Under this symmetry the superfields transform as $Q \rightarrow \Gamma Q$, $\bar{Q} \rightarrow 
\Gamma^{*}\bar{Q}$, and $V \rightarrow \Gamma V \Gamma^{\dagger}$.  Here $V$ is the vector superfield while 
$Q$ and $\bar{Q}$ are chiral superfields which transform as the fundamental and anti-fundamental 
representations of U(2$N$), and the matrix $\Gamma$ is acting on the gauge indices of $Q$ and $\bar{Q}$. We 
label this symmetry by $Z_{2 \Gamma}$. The theory is further invariant under a different discrete $Z_2$ 
symmetry which is an element of the U($2N$) $\times$ U($2N$) flavor symmetry and which is generated by 
a matrix {\bf F} of the form
\begin{equation}
\label{F}   
\pmatrix{+1 & & & & & \cr
          & \ddots & & & & \cr
        & & +1  & & & \cr
        & & &-1 & & \cr  
        & & & &\ddots & \cr
        & & & & & -1}
\end{equation}
Under this symmetry, which we denote by $Z_{2 {\rm {\bf F}}}$, the superfields $Q$ and $\bar{Q}$ transform as 
$Q \rightarrow Q {\rm {\bf F}}^{\dagger} $, $\bar{Q} \rightarrow \bar{Q} {\rm {\bf F}}^{T}$. Here the matrix 
{\bf F} acts on the flavor indices of $Q$ and $\bar{Q}$, and not the color indices. Finally, the theory is 
invariant under a $Z_{2R}$ discrete symmetry under which all bosonic components of the superfields are even 
while all fermionic components are odd. Under the combined $Z_{2 \Gamma} \times Z_{2 {\rm {\bf F}}} \times 
Z_{2 R}$ symmetry each field in any given supermultiplet is either even or odd. Specifically, for the 
components of the vector superfield $V$ of U(2$N$)
\begin{eqnarray}
A_{\mu} &=&  \pmatrix{ A_{\mu, AA} (+) & A_{\mu, AB} (-) \cr
                     A_{\mu, BA} (-) & A_{\mu, BB} (+)}
\nonumber
\\
\lambda &=&  \pmatrix{ \lambda_{AA} (-) & \lambda_{AB} (+) \cr
                     \lambda_{BA} (+) & \lambda_{BB} (-)}
\end{eqnarray}
Here $A$ and $B$ distinguish between the two U($N$) gauge groups that are contained in the original U(2$N$) 
gauge 
group. The plus and minus signs in brackets indicate whether the corresponding field is even or odd under the 
discrete symmetry. For the components of the chiral superfield $Q$
\begin{eqnarray}
\tilde{q} &=&  \pmatrix{\tilde{q}_{Aa} (+) & \tilde{q}_{Ab} (-) \cr
                        \tilde{q}_{Ba} (-) & \tilde{q}_{Bb} (+)}
\nonumber
\\
q &=&  \pmatrix{q_{Aa} (-) & q_{Ab} (+) \cr
                q_{Ba} (+) & q_{Bb} (-) }
\end{eqnarray}
Here $a$ and $b$ distinguish between the two U($N$) flavor groups that are contained in the original 
U(2$N$) 
flavor group.  The components of $\overline{Q}$ have exactly the same transformation properties as the 
corresponding components in $Q$.

We orbifold the supersymmetric U(2$N$) gauge theory down to a non-supersymmetric U($N$) $\times$ U($N$) 
daughter gauge theory by keeping in the Lagrangian only those fields invariant under the combined $Z_{2 
\Gamma} \times Z_{2 {\rm {\bf F}}} \times Z_{2 R}$ symmetry. At the same time the gauge coupling constant 
of the daughter theory is rescaled to be a factor of $\sqrt{2}$ larger than the corresponding coupling 
constant in the mother theory.  It has been shown in {\cite{fold4}} that in the large N limit the 
correlation functions of this daughter theory are equal (up to rescalings) to the corresponding correlation 
functions of the supersymmetric parent theory. This result holds to all orders in perturbation theory.

This implies that in the large $N$ limit, quadratically divergent contributions to the mass squared of the 
scalar $\tilde{q}_{Aa}$ are absent in the daughter theory. It is straightforward to verify this at one 
loop. From the couplings to the gauge bosons $A_{\mu, AA}$ of SU($N$) and U(1) we obtain quadratically 
divergent contributions $(3/32 \pi^2) g^2 \Lambda^2 (N - 1/N)$ and $(3/32 \pi^2) g^2 \Lambda^2 (1/ N)$ 
respectively. Here $g$ is the gauge coupling constant in the daughter theory and $\Lambda$ is a hard cutoff 
scale. From the scalar self-interactions that survive from the D-term of U(2$N$)  we obtain $1/32 \pi^2 g^2 
\Lambda^2 N$. Only $\lambda_{AB}$ and $\lambda_{BA}$, the off-diagonal components of the gauginos 
$\lambda$, survive after orbifolding. These contribute $- (1/8 \pi^2) g^2 \Lambda^2 N$ to the mass of 
$\tilde{q}_{Aa}$. The sum total is exactly zero, as expected from the non-renormalization theorem. What if 
we had started with SU(2$N$) instead of U(2$N$)? Then the cancellation would have been only partial, and 
the contribution to the scalar mass would have been $- (1/16 \pi^2) g^2 \Lambda^2 (1/N)$, which vanishes in 
the large $N$ limit but not otherwise.

It is important to note that in the orbifolded theory the relation between the gauge coupling constant, the 
scalar-fermion-gaugino coupling and the scalar self coupling that are crucial to this cancellation do not 
immediately follow from any symmetry principle. Therefore it is important that an ultraviolet completion 
exist that guarantees the relation between these different couplings at some ultraviolet scale. We will 
defer the problem of finding such ultraviolet completions to the next section.

Can the correspondence be extended to theories with Yukawa couplings? For certain classes of theories where 
all matter is in bifundamentals, this is straightforward.  Consider a supersymmetric
theory with an SU(2$N)_1 \times$ 
SU(2$N)_2 \times$ SU(2$N)_3$ global symmetry and matter content
$Q_{12} = (2N, \overline{2N}, 1), Q_{23} = (1, 2N, \overline{2N}), Q_{31} = (\overline{2N}, 1,
2N)$. This choice of global symmetries admits the Yukawa interaction
\begin{equation}
\lambda \; Q_{12} \; Q_{23} \; Q_{31}
\end{equation} 
in the superpotential. The Lagrangian is then invariant under a discrete $Z_{2 \Gamma}$ symmetry
where the superfields $Q_{12}$, $Q_{23}$ and $Q_{23}$ transform as $Q_{12} \rightarrow \Gamma
Q_{12} \Gamma^{\dagger}$, $Q_{23} \rightarrow \Gamma Q_{23} \Gamma^{\dagger}$ and $Q_{31} \rightarrow 
\Gamma Q_{31} \Gamma^{\dagger}$, in a notation where $\Gamma$ always acts on the SU(2$N$) index and
$\Gamma^{\dagger}$ on the SU$(\overline{2N})$ of the $Q$'s. The theory is also invariant under a 
$Z_{2R}$
discrete symmetry under which the bosonic components of each superfield are even while the
fermionic components are odd. We can obtain a daughter theory with [SU($N)_{1A} \times$
SU($N)_{1B} \times $ SU($N)_{2A} \times$ SU($N)_{2B} \times $ SU($N)_{3A} \times$ SU($N)_{3B}$]
global symmetry by projecting out of the theory those states that are odd under the combined $Z_{2
\Gamma} \times Z_{2 R}$ symmetry. Here $A$ and $B$ again distinguish between the two SU($N$)  
groups which emerge from each of the original SU(2$N$) groups. Under the action of the combined
symmetry the components of $Q_{12}$ transform as shown below. 
\begin{eqnarray} 
\tilde{q}_{12} =
\pmatrix{ \tilde{q}_{1A, 2A} (+) & \tilde{q}_{1A,2B} (-) \cr
                           \tilde{q}_{1B,2A} (-) & \tilde{q}_{1B,2B} (+)} 
\nonumber
\\
q_{12} =  \pmatrix{ q_{1A,2A} (-) & q_{1A,2B} (+) \cr
                   q_{1B,2A} (+) & q_{1B,2B} (-)}
\end{eqnarray}
The transformation of the components of $Q_{23}$ and $Q_{31}$ under the combined symmetry is identical to 
that of the corresponding components of $Q_{12}$. We also rescale $\lambda \rightarrow \sqrt{2} \lambda$ in 
the daughter theory. Then, using the methods of {\cite{fold4}} it can be shown that the correlation 
functions of the mother and daughter theories are related to all orders in perturbation theory, in exact 
analogy to the gauge theory case we studied previously.

Let us again verify the cancellation of one loop quadratically divergent contributions to the mass of the 
scalars in the daughter theory. For simplicity, we focus on corrections to the mass of $\tilde{q}_{1A, 
2A}$. Schematically, the relevant couplings are
\begin{eqnarray}
\left[\; \sqrt{2} \lambda \; \tilde{q}_{1A, 2 A} \; q_{2 A, 3 B} \; q_{3B, 1 A} 
\; + \; {\rm h.c.} \; \right] 
\; \; + \; \; \; 
\nonumber 
\\ 
2 \; \lambda^2 \; |\tilde{q}_{1A, 2 A}|^2 \; |\tilde{q}_{3 A, 1 A}|^2  +
2 \; \lambda^2 \; |\tilde{q}_{1A, 2A}|^2 \; |\tilde{q}_{2 A, 3 A}|^2 
\end{eqnarray}
It is easy to see that just as in supersymmetry, the contributions from the scalar loops cancel against the 
fermion loop so that the net one loop correction to the mass of $\tilde{q}_{1A, 2A}$ vanishes identically, 
even for small $N$. Similar cancellations extend to the masses of all other scalars in the theory. The 
correspondence implies that this cancellation goes through to all loop orders at large $N$.

\subsection{The Underlying Mechanism and Bifold Protection}\label{bifold}

In general, the class of theories to which the large $N$ orbifold correspondence applies is rather 
restricted, which would seem to limit its application to the problem of stabilizing the weak scale. 
However, in order to address the LEP paradox it is sufficient that the quadratic divergences of the SM be 
cancelled at one loop, and then again only for one specific field - the Higgs. If we can identify the 
origin of the cancellation of one loop quadratic divergences in the non-supersymmetric daughter theories 
above, it may be possible to apply the same underlying principles to construct much larger classes of 
theories which are radiatively stable at one loop.

What then underlies the cancellation of one loop quadratic divergences in the examples we have considered? 
The key observation is that in each case the scalar mass in the mother theory enjoys {\bf bifold 
protection}. Consider one loop quadratically divergent corrections to the scalar mass in the mother theory. 
For any given graph the states running in the loop each carry two large $N$ indices. One of these indices, 
which we label $`i'$, is summed over from 1 to $2N$, while the other index is unsummed. Consider the 
contribution arising from bosons running in the loop, with the summed index $i$ running from 1 to $N$.  
This can be thought of as being cancelled {\bf either} by the fermion loop with $i$ again running from 1 to 
$N$ {\bf or} by the fermion loop with $i$ instead running from $N + 1$ to $2N$. The first cancellation is 
an immediate consequence of supersymmetry. The second follows from the combination of supersymmetry and 
additional global, gauge or discrete symmetries that these theories possess. Then by projecting out of the 
theory those bosons with index $i$ running from $N + 1$ through $2N$, and also those fermions with index 
$i$ running from 1 through $N$, the cancellation still goes through. This explains the absence of one loop 
quadratic divergences to the scalar mass in the daughter theory.

Based on this observation, we are now in a position to outline a set of procedures which suitably extend 
the particle content and vertices of a theory so as to cancel the leading one loop quadratic divergence to 
the mass of a scalar arising from a specific interaction. The `rules' below apply in most simple cases,
including those we will be considering.
\begin{itemize}
\item
Supersymmetrize.
\item
In the relevant graphs identify an index as being summed over from 1 to $N$. Then extend the particle 
content and gauge, global or discrete symmetries of the theory so that this index runs from 1 to 2$N$, 
while the vertices in each graph otherwise remain the same. For the cases of SU($N$) gauge 
interactions and Yukawa interactions, this can always be done in such a way that the scalar mass parameter 
in the resulting theory enjoys bifold protection, and is invariant under $Z_{2 \Gamma}$ and $Z_{2 R}$ 
symmetries.
\item
Project out states odd under the combined $Z_{2 \Gamma} \times Z_{2R}$ symmetry. The resulting daughter 
theory is free of one loop quadratic divergences, up to potential (1/$N$) corrections. 
\end{itemize}
When applied to SU(N) gauge interactions, or to Yukawa interactions, an ultraviolet completion can 
always be found for the daughter theory that is consistent with this cancellation. We will see how
to construct such ultra-violet completions in the next section.

We now provide an example of how to apply these rules.
Consider a theory consisting of a scalar singlet $S$ that has a Yukawa coupling to chiral 
fermions $Q_i$ and $\bar{Q}_i$ which transform as the fundamental and anti-fundamental representations of a 
global U($N$) symmetry. Here the index $i$ runs from 1 to $N$. The Yukawa coupling takes the form
\begin{equation}
\lambda \; S \; Q_i \; \overline{Q}_i
\end{equation} 
We wish to extend this theory so as to cancel quadratic divergences to the scalar mass from this Yukawa 
interaction. We first supersymmetrize so that $S$, $Q_i$ and $\bar{Q}_i$ are all promoted to chiral 
superfields and the Yukawa interaction above is now in the superpotential. We identify $i$ as the large $N$ 
index since it is summed over in the loop which contributes to the mass of $S$. We therefore promote the 
global U($N$) symmetry to a global U($2N$) symmetry by adding extra $Q$'s and $\bar{Q}$'s to the theory so 
that the index $i$ now runs from 1 to $2N$, while the Yukawa coupling above has exactly the same form as 
above. In the resulting theory the mass of the scalar $S$ clearly enjoys bifold protection. The theory also 
possesses a $Z_{2 \Gamma}$ symmetry under which the singlet $S$ is invariant while $Q \rightarrow - \Gamma 
Q$, $\bar{Q} \rightarrow - \Gamma^{*}\bar{Q}$, and a $Z_{2 R}$ symmetry under which all bosonic fields are 
even while all fermionic fields are odd. If we project out all fields odd under the combined $Z_{2 \Gamma} 
\times Z_{2 R}$ symmetry, it is straightforward to verify that in the daughter theory quadratically 
divergent contributions to the mass of $S$ vanish even though the theory is not supersymmetric.  However, 
note that quadratically divergent contributions to the mass of the scalars in $Q$ and $\bar{Q}$ in the 
daughter theory, while large $N$ suppressed, do not in fact cancel at all. This will feed into the mass of 
$S$ at one higher loop order, and therefore the procedure we have outlined to protect the mass of the 
scalar $S$ does not extend beyond one loop. However, as we have explained, this is perfectly sufficient to 
address the LEP paradox.

\section{Application to the Standard Model} \label{applications}

In this section we apply these ideas to the problem of stabilizing the weak scale. We limit ourselves to 
finding appropriate orbifolds and their ultraviolet completions, while postponing the discussion of 
completely realistic models to the next section. The Higgs mass parameter in the SM receives one loop 
quadratically divergent contributions from gauge, Yukawa and quartic self-interactions. Of these the 
contribution from the top Yukawa coupling is numerically the most significant by about an order of 
magnitude, and we therefore consider it first. We then go on to consider a model where the dominant part of 
the one loop quadratic divergence from the gauge interactions is cancelled.

\subsection{The Top Yukawa Coupling}
\label{top}

\noindent
{\bf Choice of Orbifold}
\medskip

After supersymmetrization the top Yukawa interaction has the form
\begin{equation}
\lambda_t \left(3,2 \right)_{Q_3} \left(1,2 \right)_{H_U} \left(\overline{3},1 \right)_{U_3}  
\end{equation}
in the superpotential. Here $Q_3$ represents the third generation SU(2) doublet containing the top and 
bottom quarks, $H_U$ the up-type Higgs and $U_3$ the SU(2) singlet (anti)top-quark. If we treat both SU(2) 
indices $i$ and SU(3) indices $\alpha$ as large $N$ indices, in t'Hooft double line notation the top quark 
contribution to the Higgs mass parameter takes the form shown in Figure~(\ref{largeN-top}). 
\begin{figure}
\begin{center}
\includegraphics[width=0.7\columnwidth]{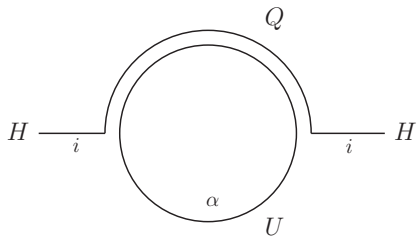}
\end{center}
\caption{The top loop diagram for the Higgs mass in double line notation. In order to give the Higgs bifold protection SU(3) color, represented by the index $\alpha$, may be extended to~SU(6) or to~SU(3$)\times$ SU(3$)\times Z_2$.} 
\label{largeN-top}
\end{figure}
From the figure, it
is clear that it is the SU(3) indices $\alpha$ which are being summed over. In order to obtain a 
theory where the Higgs mass enjoys bifold 
protection we must double this sum. This can be done in either of two ways, which have somewhat
different phenomenology.
\begin{itemize}
\item
Extend the gauge symmetry from SU(3) to SU(6). This is the approach we shall follow for the rest
of this section.
\item
Extend the gauge symmetry from SU(3) to [SU(3) $\times$ SU(3)], with a discrete symmetry interchanging
the two SU(3) gauge groups. We will consider this approach in section~\ref{realistic}.
\end{itemize} 
After extending the SU(3) color gauge symmetry of the SM to an SU(6) gauge symmetry, the top Yukawa 
coupling has the form
\begin{equation}
\lambda_t \left(6,2 \right)_{{Q}_{3T}} \left(1,2 \right)_{H_U} \left(\overline{6},1 
\right)_{{U}_{3T}}
\end{equation}
Here the field ${{Q}_{3T}}$ contains not only ${Q_3}$ of the SM but also exotic fields charged under 
SU(2$)_{\rm L}$ and U(1$)_{\rm Y}$ but not under SM color.  Similarly ${{U}_{3T}}$ contains not only 
${U_3}$ of the SM but also exotic fields charged under U(1$)_{\rm Y}$ but not under SM color. We refer to 
these new fields as the `folded partners' (or `F-partners' for short) of the corresponding MSSM fields. 
Now 
the theory is invariant under a $Z_{2 \Gamma}$ symmetry under which ${{Q}_{3T}} \rightarrow - \Gamma 
{{Q}_{3T}}$, ${{U}_{3T}} \rightarrow - \Gamma^{*} {{U}_{3T}}$, $V_6 \rightarrow \Gamma V_6 
\Gamma^{\dagger}$. Here $V_6$ is the vector superfield corresponding to the SU(6) gauge group. The form of 
the matrix $\Gamma$ is as shown in Eq. (\ref{Gamma}). We temporarily defer the question of how this 
symmetry is extended to the other fields in the MSSM. The theory also possesses a $Z_{2R}$ symmetry under 
which all fermionic fields are odd and all bosonic fields even.

Now consider the transformation properties of the various fields under the combined $Z_{2 \Gamma}
\times Z_{2R}$ symmetry. 
\begin{equation}
\tilde{q}_{3T} =  \pmatrix{\tilde{q}_{ \alpha} (-)\cr
                      \tilde{q}_{ \beta} (+)}
\;\;\;\;
q_{3T} =  \pmatrix{q_{ \alpha} (+) \cr
              q_{ \beta} (-)}
\end{equation}
Here $\alpha$ and $\beta$ distinguish between the two SU(3) subgroups of SU(6)
which are left unbroken
under this operation. Similarly
\begin{equation}
\tilde{u}_{3T} =  \pmatrix{\tilde{u}_{ \alpha} (-)\cr
                      \tilde{u}_{ \beta} (+)}
\;\;\;\;
u_{3T} =  \pmatrix{u_{ \alpha} (+) \cr
              u_{ \beta} (-)}
\end{equation}
while the scalar and fermion components of $H_U$ are even and odd respectively.
\begin{equation}
H_U = \left( h_u (+), \tilde{h}_u (-) \right) 
\end{equation}
After orbifolding out the odd states, consider the quadratically divergent contributions to the mass 
parameter of the up-type Higgs field. 
The relevant interactions have the form
\begin{eqnarray}
\label{higgstop}
\left[\; \lambda_t \; h_{u} \; q_{ \alpha} \; u_{ \alpha} \; + \; {\rm h.c.} \; \right]
\; \; + \; \; \;
\nonumber
\\
\lambda_t^2 \; |\tilde{q}_{ \beta} h_u|^2  +
\lambda_t^2 \; |\tilde{u}_{ \beta}|^2 \; |h_u|^2
\end{eqnarray}
Then quadratically divergent contributions from scalar loops cancel against those from
fermion loops. Note, however that the scalar fields responsible for this cancellation are 
not charged under SM color, but under a different, hidden color group. 
 
What about quadratically divergent contributions to the masses of the F-squarks $\tilde{q}_{
\beta}$ and $\tilde{u}_{ \beta}$? It is easy to see that these do not cancel, because these
fields do not have any couplings to fermions in the daughter theory.  This implies that there
will be quadratically divergent contributions to the mass of the Higgs at two loops. This is an
illustration of the fact that for general orbifolds the daughter theory does not possess any
symmetry that can guarantee radiative stability of the parameters to all orders. For this reason
it is important that the daughter theory possess an ultraviolet completion that can set the
values of the parameters at the high scale. 

\medskip
\noindent
{\bf An Ultraviolet Completion}
\medskip

We now outline an ultraviolet completion that sets the couplings of the Higgs field in the low energy
effective 
theory to their 
folded-supersymmetric values. Consider a five-dimensional supersymmetric theory with an extra dimension of 
radius $R$ compactified on $S^1/Z_2$, with branes at the orbifold fixed points. The locations of the branes 
are at $y = 0$ and $y = \pi R$, where $y$ denotes the coordinates of points in the fifth dimension.  The 
gauge symmetry is SU(6$) \times$ SU(2$) \times$ U(1), and all gauge fields live in the bulk of the higher 
dimensional space. The SU(6) gauge symmetry is broken to SU(3$) \times$ SU(3$) \times$ U(1) by boundary 
conditions{\cite{stringorbifolds,Kawamura,Altarelli,Hall}}. At the same time supersymmetry is broken by the 
Scherk-Schwarz mechanism{\cite{S&S,P&Q,APQ,BHN}} so that while physics on each brane respects a 
(different) four dimensional $\mathcal{N}=1$ supersymmetry, below the compactification scale supersymmetry 
is 
completely broken. The Higgs fields $H_U$ and $H_D$ are localized on the brane where SU(6) is preserved. 
However all matter fields emerge from hypermultiplets which live in the bulk of the space. To specify the 
boundary conditions to be satisfied by bulk fields we need to know their transformation properties under 
reflections about $y=0$, which we denote by $Z$. In addition, we also need to specify either their 
transformation properties under translations by $2 \pi R$, which we denote by $T$, or their transformation 
properties under reflections about $\pi R$, which we denote by $Z'$. $T$ and $Z'$ are related by $Z' = T \; 
Z$. We choose to describe the boundary conditions satisfied by the various fields in terms of $Z$ 
and~$Z'$.

A supersymmetric gauge multiplet $\hat{V}$ in five dimensions consists of $A_M, \lambda,
\lambda'$ and $\sigma$. From the four dimensional viewpoint the five dimensional theory has 
$\mathcal{N}=2$
supersymmetry. Under the action of $Z$ this $\mathcal{N}=2$ supersymmetry is broken to $\mathcal{N}=1$ 
supersymmetry.
The five dimensional multiplet can be broken up into four dimensional $\mathcal{N}=1$ supermultiplets as
$\hat{V} = \left(V, \Sigma \right)$ where $V$ consists of $\left(A_{\mu}, \lambda \right)$ and
$\Sigma$ of $\left( \sigma + i A_5, \lambda' \right)$. $V$ and $\Sigma$ must necessarily have
different transformation properties under $Z$. Similarly, under the action of $Z'$ the four
dimensional $\mathcal{N}=2$ supersymmetry is also broken to $\mathcal{N}=1$ supersymmetry. However, since 
we are
interested in Scherk-Schwarz supersymmetry breaking this $\mathcal{N}=1$ supersymmetry must be different
from that which survives the operation $Z$. An alternative decomposition of the five dimensional
multiplet into four dimensional $\mathcal{N}=1$ multiplets is $\hat{V} = \left(V', \Sigma' \right)$ where
$V'$ consists of $\left(A_{\mu}, \lambda' \right)$ and $\Sigma'$ of $\left( \sigma + i A_5, -
\lambda \right)$.  This new decomposition is related to the first one by an SU(2$)_R$ rotation.
We require that $V'$ and $\Sigma'$ have different transformation properties under $Z'$. Then the
combined action of $Z$ and $Z'$ breaks supersymmetry completely. The fields which have zero modes
in the low energy theory are those which are even under the action of both $Z$ and~$Z'$.

A hypermultiplet $\hat{Q}$ in five dimensions consists of bosonic fields $\tilde{q}$ and
$\tilde{q}^c$ and fermionic fields $q$ and $q^c$. The hypermultiplet can be decomposed into four
dimensional $\mathcal{N}=1$ superfields. Then $\hat{Q}$ breaks up into $(Q, Q^c)$ where $Q = (\tilde{q}, 
q)$
and $Q^c =(\tilde{q}^c, q^c)$. Since $Q$ and $Q^c$ have different transformation properties
under $Z$, the four dimensional $\mathcal{N}=2$ supersymmetry of the system is broken to $\mathcal{N}=1$.  
An
alternative decomposition of the five dimensional hypermultiplet into four dimensional $\mathcal{N}=1$
superfields is $\hat{Q} = (Q', Q'^c)$ where $Q' = (\tilde{q}^{*c}, q)$ and $Q'^c = (-\tilde{q}^*,
q^c)$. This new decomposition of $\hat{Q}$ is related to the first by the same SU(2$)_R$ rotation as
in the case of the gauge supermultiplet. To break supersymmetry we require that $Q'$ and $Q'^c$
necessarily have different transformation properties under $Z'$. Although individually each of
$Z$ and $Z'$ preserve one four dimensional $\mathcal{N}=1$ supersymmetry, their collective action breaks
supersymmetry completely.

In order to break SU(6) to SU(3$) \times$ SU(3$) \times$ U(1) we must impose suitable boundary
conditions. We choose $Z$ to leave SU(6) unbroken while $Z'$ breaks SU(6). Therefore, if we
denote the five dimensional SU(6) gauge multiplet by $\hat{V}_6$, then under the action of $Z$,
$V_6$ is even and $\Sigma_6$ odd. However, under the action of $Z'$
\begin{equation}
Z':\qquad
V'_6 \rightarrow \Gamma
V'_6 \Gamma^{\dagger}   
\qquad
\Sigma'_6 \rightarrow - \Gamma \Sigma'_6 \Gamma^{\dagger}. 
\end{equation}
Then
the gauge bosons of SU(3$) \times$ SU(3$) \times$ U(1), together with the fields in $\lambda_6$
which have the quantum numbers of SU(6)/[SU(3$) \times$ SU(3$) \times$ U(1)], are present in the
low energy theory whereas all other fields in $\hat{V}_6$ are projected out. However we wish to
leave SU(2$) \times$ U(1) of the SM unbroken, so for these vector multiplets we simply keep $V$ even
under $Z$ and $V'$ even under $Z'$ while projecting out $\Sigma$ and $\Sigma'$. Only the gauge
bosons of SU(2$) \times$ U(1) are then present in the low energy spectrum.

We now turn our attention to the boundary conditions on the matter hypermultiplets involved in the
top Yukawa coupling. Introduce into the bulk a hypermultiplet ${\hat{Q}}_{3T}$ which transforms as
(6,2) under SU(6$) \times$ SU(2) and has hypercharge (1/3). Under $Z$ $\hat{Q}_{3T}$ breaks up into
(${Q}_{3T}, {Q}_{3T}^c$) where ${Q}_{3T} = (\tilde{q}_T, q_T)$ is even while $\hat{Q}_{3T}^c =
(\tilde{q}_T^c, q_T^c)$ is odd.  Under $Z'$ we have 
\begin{equation}
Z':\qquad
{Q}'_{3T} \rightarrow \Gamma {Q}'_{3T}, \qquad
{{Q}_{3T}}'^c \rightarrow - \Gamma^{*} {{Q}_{3T}}'^c.
\end{equation} 
Then the fields which have zero modes are the
fermion $q_{\alpha}$ and the scalar $\tilde{q}_{\beta}$. To obtain $U_3$ introduce into the bulk a
hypermultiplet $\hat{U}_{3T}$ which transforms as $(\overline{6}, 1)$ under SU(6$) \times$ SU(2)
and has hypercharge -(4/3). Under $Z$ $\hat{U}_{3T}$ breaks up into (${U}_{3T}, {U}_{3T}^c$) where
${U}_{3T} = (\tilde{u}_T, u_T)$ is even while ${U}_{3T}^c = (\tilde{u}_T^c, u_T^c)$ is odd.  Under
$Z'$ we have 
\begin{equation}
Z':\qquad
U_{3T}' \rightarrow \Gamma^{*} U_{3T}', \qquad
{U'_{3T}}^c \rightarrow - \Gamma
{U'_{3T}}^c. 
\end{equation}
Then the fields which have zero modes are the fermion $u_{\alpha}$ and the scalar
$\tilde{u}_{\beta}$.

Now consider the top Yukawa coupling written on the brane at $y=0$.
\begin{equation}
\lambda_t \left(6,2 \right)_{{Q}_{3T}} \left(1,2 \right)_{H_U} 
\left(\overline{6},1 \right)_{{U}_{3T}}
\end{equation}
In the four dimensional effective theory obtained after integrating out the Kaluza-Klein modes the 
couplings of the Higgs scalar have exactly the form of Eq.(\ref{higgstop}), and so there is no one loop 
contribution to the Higgs mass parameter from the light fields.

One may worry that the Kaluza-Klein tower, being non-supersymmetric, will contribute a large radiative 
correction to the Higgs mass.  In the appendix it is shown that there is also no contribution from the 
Kaluza-Klein states. This is because the Kaluza-Klein tower has equal numbers of bosonic and fermionic 
states at every level, as depicted schematically in Figure~(\ref{fig-spectrum}), and the couplings of these 
states to the Higgs are related in such a way as to exactly guarantee cancellation at every level. Note 
however that the cancellation is occurring between states which do not have the same charge under SU(3) 
color.
\begin{figure}
\begin{center}
\includegraphics[width=0.7\columnwidth]{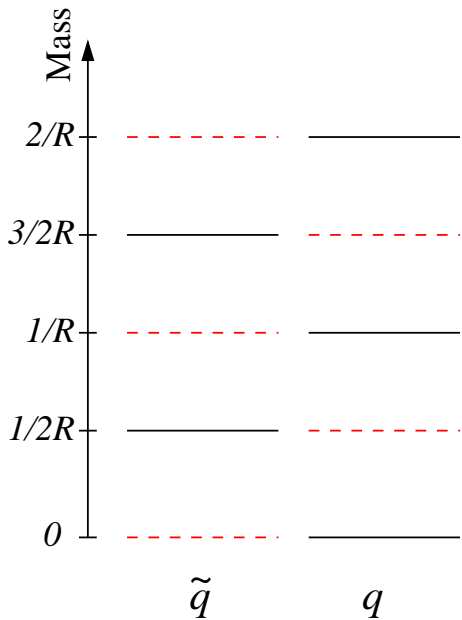}
\end{center}
\caption{The Kaluza-Klein tower in theories with Scherk-Schwarz SUSY breaking admits non-degenerate 
fermions and 
bosons. However in folded-supersymmetric theory each such tower is complemented by another tower yielding a 
degenerate spectrum. This allows for a complete cancellation of radiative corrections to the Higgs mass 
at one loop.} \label{fig-spectrum}
\end{figure}

\subsection{SU(2) Gauge Interactions} 
\label{gauge}

\noindent
{\bf Choice of Orbifold}
\medskip

We now consider how to cancel the dominant one loop quadratically divergent contributions to the Higgs mass 
from SU(2) gauge interactions. If we treat SU(2) indices $i$ as large $N$ indices, in t'Hooft double line 
notation the gauge contribution to the Higgs mass parameter takes the form shown 
in Figure~(\ref{largeN-gauge}). 
\begin{figure}
\begin{center}
\includegraphics[width=0.5\columnwidth]{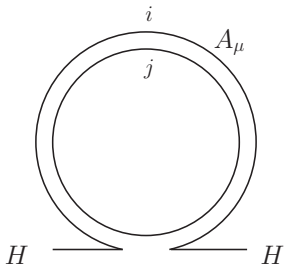}
\end{center}
\caption{The gauge loop contribution to the Higgs mass in double line notation. In order to give the Higgs bifold protection SU(2) may be extended to SU(4).} \label{largeN-gauge}
\end{figure}
From the diagram it is clear that it is SU(2) indices which are being summed over in the loop. Therefore, in order to obtain a theory where the Higgs mass enjoys bifold protection, we must supersymmetrize and double the sum over SU(2) indices. One way of doubling the sum is to extend the SM gauge structure from SU(3$) \times$ SU(2$) \times$ U(1) to SU(3$) \times$ SU(4$) \times$ U(1). The up and down-type Higgs fields, ${H}_U$ and ${H}_D$ then transform as $4$ and $\overline{4}$ under the SU(4) symmetry. The resulting theory possesses a $Z_{2 \Gamma}$ symmetry under which ${H}_U \rightarrow \Gamma {H}_U$, 
${H}_D \rightarrow \Gamma^{*} {H}_D$. Also $V_4 \rightarrow \Gamma_4 V_4 \Gamma_4^{\dagger}$. As before the theory is 
invariant under a $Z_{2R}$ symmetry under which all bosonic fields are even and all fermionic fields odd. 
We now consider the transformation properties of the various fields under the combined $Z_{2 \Gamma} \times 
Z_{2R}$ symmetry.  For the components of the field $H_U$
\begin{equation}
{h_U} =  \pmatrix{h_{UA} (+)\cr
                  h_{UB} (-)}
\;\;\;\;
\tilde{h}_U =  \pmatrix{\tilde{h}_{UA} (-)\cr
                        \tilde{h}_{UB} (+)}
\end{equation}
For the components of $V_4$,
\begin{eqnarray}
A_{\mu} &=&  \pmatrix{ A_{\mu, AA} (+) & A_{\mu, AB} (-) \cr
                     A_{\mu, BA} (-) & A_{\mu, BB} (+)}
\nonumber
\\
\lambda &=&  \pmatrix{ \lambda_{AA} (-) & \lambda_{AB} (+) \cr
                     \lambda_{BA} (+) & \lambda_{BB} (-)}
\end{eqnarray}
Here $A$ and $B$ distinguish between the two SU(2) subgroups of SU(4). We now project out states
odd under $Z_{2 \Gamma} \times Z_{2R}$. The gauge symmetry is then broken down to SU(2$) 
\times$ SU(2$) \times$ U(1). 
Let us consider contributions from this sector 
to the mass of the Higgs scalar $h_{UA}$. Schematically, the
relevant interactions are 
\begin{eqnarray}
\label{gaugehiggs}
\left|\left( \partial_{\mu} - ig A_{\mu, AA} - i\frac{g}{\sqrt{2}} A_{\mu, D} \right) h_{UA}
\right|^2 +
\nonumber \\
\left[ i \sqrt{2} g \; h_{UA}  \lambda_{AB} \tilde{h}_{UB} + {\rm h.c.} \right] + 
\left[ \; {\rm D-terms}  \; \right]
\end{eqnarray}
where $A_{\mu, D}$ represents the gauge boson of the unbroken diagonal U(1). The SU(2) gauge
interactions contribute $9/64 \pi^2 g^2 \Lambda^2$ to the mass of $h_{UA}$, while from the scalar
self-interactions that survive in the SU(2) D-term we obtain $3/64 \pi^2 g^2 \Lambda^2 $.  The
off-diagonal components of the SU(4) gauginos $\lambda_{AB}$ and $\lambda_{BA}$ contribute $- 1/4
\pi^2 g^2 \Lambda^2 $, and finally $A_{\mu, D}$ and its D-term together contribute $1/32 \pi^2
g^2 \Lambda^2$.  The sum total is $- 1/32 \pi^2 g^2 \Lambda^2$, and so the cancellation is
incomplete. This is because we started from SU(4) and not from U(4). Nevertheless, since the
naive SM estimate of the contribution to the Higgs mass from SU(2) gauge loops is $9/64 \pi^2 g^2
\Lambda^2$, this still represents an improvement over the SM by a factor of about 5 or so.
However, the fact that the result is quadratically divergent means that 
whether this improvement is significant or not depends on whether a
ultraviolet completion exists that is naturally consistent with this cancellation.

\medskip
\noindent
{\bf An Ultraviolet Completion}
\medskip

We now outline such an ultraviolet completion. As before we consider a five-dimensional supersymmetric 
theory with an extra dimension of radius $R$ compactified on $S^1/Z_2$, with branes at the orbifold fixed 
points. As before the branes are at $y = 0$ and $y = \pi R$. The gauge symmetry is SU(3$) \times$ SU(4$) 
\times$ U(1), and all gauge fields live in the bulk of the higher dimensional space. This time it is the 
SU(4) gauge symmetry which is broken to SU(2$) \times$ SU(2$) \times$ U(1) by boundary conditions.  As 
before supersymmetry is also broken by the Scherk-Schwarz mechanism, so that while physics on each brane 
respects a (different) four dimensional $\mathcal{N}=1$ supersymmetry, below the compactification scale 
supersymmetry is completely broken.  The Higgs fields $H_U$ and $H_D$ are localized on the brane where 
SU(4) is preserved.

In order to break SU(4) to SU(2$) \times$ SU(2$) \times$ U(1) we must impose suitable boundary conditions. 
We choose $Z$ to leave SU(4) unbroken while $Z'$ breaks SU(4). Then if we denote the five dimensional SU(4) 
gauge multiplet by $\hat{V}_4$, then under the action of $Z$, $V_4$ is even and $\Sigma_4$ odd. However, 
under the action of $Z'$, $V'_4 \rightarrow \Gamma V'_4 \Gamma^{\dagger}$ while $\Sigma'_4 \rightarrow - 
\Gamma \Sigma'_4 \Gamma^{\dagger}$. Then the gauge bosons of SU(2$) \times$ SU(2$) \times$ U(1), together 
with the fields in $\lambda_4$ which have the quantum numbers of SU(4)/[SU(2) $\times$ SU(2) $\times$ 
U(1)], are the only ones with zero modes. If we now consider the couplings of the Higgs fields $H_U$ and 
$H_D$ to the zero modes of $\hat{V}_4$, they now have exactly the form of Eq. (\ref{gaugehiggs}), with the 
remaining quadratic divergence from the U(1) cutoff by the Kaluza-Klein modes. The net contribution to the 
Higgs mass parameter in this theory is calculated in the appendix. It is non-zero but finite and about a 
factor of 20 smaller than the naive SM estimate of $9/64 \pi^2 g^2 \Lambda^2$ when $\Lambda$ is 
replaced by $1/R$. Since we wish to leave SU(3$) \times$ U(1$)_{\rm Y}$ of the SM unbroken, for these 
vector multiplets we simply keep $V$ even under $Z$ and $V'$ even under $Z'$ while projecting out $\Sigma$ 
and $\Sigma'$. Only the gauge bosons of SU(3$) \times$ U(1$)_{\rm Y}$ are then present in the low energy 
spectrum. It is possible to construct a completely realistic model along these lines but we leave this for 
future work.

\section{A Realistic Model}
\label{realistic}

We now construct a realistic model based on the tools we have developed in the last two sections. In this 
example, quadratically divergent contributions to the SM Higgs mass parameter from the SU(2)  and U(1) 
gauge bosons are cancelled by the corresponding gauginos, just as in the MSSM. However, the one loop 
contributions to the Higgs mass from the top loop are cancelled by particles with no charge under SM color, 
giving rise to a very distinct and exciting phenomenology. The model is similar to the corresponding five 
dimensional model in section~\ref{top}.  The major difference is that in this model the bulk SU(6) gauge 
symmetry is replaced by a bulk SU(3$) \times$ SU(3) gauge symmetry with a $Z_2$ interchange symmetry that 
links the particle content and coupling constants of the two SU(3) gauge interactions. The SU(3) $\times$ 
SU(3) $\times Z_2$ symmetry is sufficient to ensure that the Higgs mass parameter enjoys bifold protection 
from Yukawa interactions, which allows the crucial cancellation to go through just as in the SU(6) model.

Once again we begin with a five-dimensional supersymmetric theory. The extra dimension, which
has radius $R$, is compactified on $S^1/Z_2$, and there are branes at the orbifold fixed points
$y = 0$ and $y = \pi R$. The gauge symmetry is now [SU(3$)_A \times$ SU(3$)_B] \times$
SU(2$)_{\rm L} \times$ U(1$)_{\rm Y}$, and as before all gauge fields live in the bulk of the
higher dimensional space.  While the SU(3$)_A$ gauge group corresponds to the familiar SM color,
SU(3$)_B$ corresponds to a mirror color gauge group. The remaining SU(2$)_{\rm L}$ and
U(1$)_{\rm Y}$ give rise to the SM weak and hypercharge interactions. All matter fields
arise from hypermultiplets living in the five dimensional bulk. There is a discrete $Z_2$
symmetry in the bulk that interchanges the vector superfields of the two SU(3) gauge groups,
but which acts trivially on SU(2$)_{\rm L}$ and U(1$)_{\rm Y}$ vector superfields. We label this
interchange symmetry by $Z_{AB}$.  The bulk hypermultiplets 
from which the SM quarks and their F-spartners emerge are
\begin{eqnarray}
\hat{Q}_{i A} \; \; (3,1,2,{1}/{6})  && 
\hat{Q}_{i B} \; \;  (1,3,2,{1}/{6})  \nonumber \\
\hat{U}_{i A} \; \; (\bar{3},1,1,-{2}/{3}) && 
\hat{U}_{i B} \; \;  (1,\bar{3},1,-{2}/{3})  \nonumber \\ 
\hat{D}_{i A} \; \; (\bar{3},1,1,{1}/{3})  && 
\hat{D}_{i B} \; \;  (1,\bar{3},1,{1}/{3})  
\end{eqnarray} 
where the index $A$ denotes the SM fields and $B$ their F-partners. The index $i$, which
runs from 1 to 3 labels the different SM generations.  The numbers in brackets indicate the
quantum numbers of the various fields under SU(3$)_{A} \times$ SU(3$)_{B} \times$ SU(2$)_{\rm L} \times$ 
U(1$)_{\rm
Y}$. Under the bulk $Z_{AB}$ interchange symmetry the indices $A$ and $B$ are interchanged. The
SM leptons and their F-spartners emerge from the bulk hypermultiplets below.
\begin{eqnarray} 
\hat{L}_{i A} \; \;  (1,1,2,-{1}/{2})  && \hat{L}_{i B} \; \;  (1,1,2,-{1}/{2}) \nonumber \\
\hat{E}_{i A} \; \; (1,1,1,1) && \hat{E}_{i B} \; \; (1,1,1,1) 
\end{eqnarray} 
Note that $\hat{L}_{i A}$ and $\hat{L}_{i B}$ have exactly the same gauge charges, as do
$\hat{E}_{i A}$ and $\hat{E}_{i B}$. Once again, under the bulk $Z_{AB}$ interchange symmetry
the indices $A$ and $B$ are interchanged.  The boundary conditions on the bulk hypermultiplets 
are chosen
to break both supersymmetry and the discrete $Z_{AB}$ symmetry. Specifically, we choose boundary 
conditions so that only the SM fields and their F-spartners are light.
\begin{itemize}
\item 
Of the fields $\hat{Q}_{i A}$, $\hat{U}_{i A}$, $\hat{D}_{i A}$, $\hat{L}_{i A}$ and 
$\hat{E}_{i A}$ only the 
fermions have zero modes, and
\item 
of the fields $\hat{Q}_{i B}$,  $\hat{U}_{i B}$,  $\hat{D}_{i B}$,  $\hat{L}_{i B}$ and
$\hat{E}_{i B}$ only the bosons have zero modes. 
\end{itemize} 
This is realized in the following way. When written in terms of $\mathcal{N}=1$ superfields $\hat{Q}_{i
A}$ can be decomposed into $(Q_{i A}, Q_{i A}^c)$ or into $(Q'_{i A}, Q_{i A}'^c)$.  Under the
action of $Z$, $Q_{i A}$ is even while $Q_{i A}^c$ is odd and under the action of $Z'$, $Q'_{i
A}$ is even while $Q_{i A}'^c$ is odd. These boundary conditions project out a zero mode fermion
but no corresponding light scalar. Zero mode fermions can be obtained from $\hat{U}_{i A},
\hat{D}_{i A}, \hat{L}_{i A}$ and $\hat{E}_{i A}$ by applying exactly the same boundary
conditions. 

What about the mirror fields? When written in terms of $\mathcal{N}=1$ superfields $\hat{Q}_{i B}$ can be 
decomposed into $(Q_{i B}, Q_{i B}^c)$ or into $(Q'_{i B}, Q_{i B}'^c)$.  Under the action of $Z$, $Q_{i 
B}$ is even while $Q_{i B}^c$ is odd, and under the action of $Z'$, $Q'_{i B}$ is odd while $Q_{i B}'^c$ is 
even. These boundary conditions project out a zero mode scalar but no corresponding light fermion.  Zero 
mode scalars can be obtained from $\hat{U}_{i B}$, $\hat{D}_{i B}$, $\hat{L}_{i B}$ and $\hat{E}_{i B}$ by 
applying exactly the same boundary conditions. Note that the symmetry $Z_{AB}$ is broken by the boundary 
conditions at $y = \pi R$, but not at $y=0$. This choice of brane and bulk fields implies the absence of 
mixed U(1) and gravitational anomalies anywhere in the space. Then Fayet-Iliapoulos terms are not 
radiatively generated at the boundaries~{\cite{FIterms1,FIterms2}}.
 
The MSSM Higgs fields are localized on the brane at $y =0$. We extend the $Z_2$ interchange
symmetry of the bulk to this brane. Then the Higgs couples with equal strength to both SM fields
and mirror fields, and the top Yukawa coupling has the form
\begin{equation}
W=\delta\left( y \right) \lambda_t \left[ Q_{3 A} H_U U_{3 A} +
Q_{ 3 B} H_U U_{3 B} \right]
\end{equation}
Notice that the SU(3) $\times$ SU(3) $\times Z_2$ symmetry tightly constrains the form of the top Yukawa 
coupling. In particular, this interaction takes exactly the same SU(6) symmetric form as in the 
corresponding theory in the previous section. This ensures that the Higgs mass parameter enjoys bifold 
protection from radiative corrections arising from the top Yukawa. The interactions of the Higgs in the 
four dimensional effective theory again have the folded-supersymmetric form of Eq~(\ref{higgstop}). As 
before, the one-loop contribution to the Higgs mass parameter from the top loop is cancelled by the mirror 
stops. As shown in the appendix this cancellation is not restricted to the zero-modes but persists all the 
way up the Kaluza-Klein tower and is guaranteed by a combination of supersymmetry and the discrete 
symmetry. Therefore the top Yukawa coupling does not contribute to the Higgs mass at one loop.

In this theory the top Yukawa coupling, which is required to be of order one, is volume suppressed.
This implies that the cutoff $\Lambda$ of this theory cannot be much larger than inverse of the
compactification scale, $\Lambda \lesssim 4 R^{-1}$. This leads to a potential problem. Kinetic
terms of the form $\int d^4 \theta \; Q_{3 \alpha}^{'\dagger} e^{V} Q'_{3 \alpha}$ localized on 
the
brane at $y = \pi R$ which do not respect the $Z_{AB}$ symmetry may affect the cancellation. However, this 
difficulty can be
avoided by imposing an additional symmetry on the theory. In the bulk the theory possesses a
discrete charge conjugation symmetry under which the SM matter fields are interchanged with their
corresponding charge conjugate fields in the mirror sector. This takes the form
\begin{eqnarray}
Q_{iA}' \leftrightarrow Q_{iB}'^c && Q_{iB}' \leftrightarrow - Q_{iA}'^c \nonumber \\
U_{iA}' \leftrightarrow U_{iB}'^c && U_{iB}' \leftrightarrow - U_{iA}'^c \nonumber \\
D_{iA}' \leftrightarrow D_{iB}'^c && D_{iB}' \leftrightarrow - D_{iA}'^c \nonumber \\
L_{iA}' \leftrightarrow L_{iB}'^c && L_{iB}' \leftrightarrow - L_{iA}'^c \nonumber \\
E_{iA}' \leftrightarrow E_{iB}'^c && E_{iB}' \leftrightarrow - E_{iA}'^c
\end{eqnarray}
The vector superfields of SM color are also to be interchanged with their charge conjugates in
the the mirror color sector while the vector superfields of SU(2$)_{\rm L}$ and U(1$)_{\rm Y}$
simply transform into their charge conjugates. We label this discrete symmetry by $Z'_{AB}$.  
Although the boundary conditions break this discrete symmetry on the brane at $y =0$, this
symmetry can be consistently extended to the brane at $ y = \pi R$.  This provides a restriction
on the form of the brane-localized kinetic terms at $ y = \pi R$ so that the cancellation of all
one loop corrections to the Higgs mass parameter from the top sector continues to hold.

The zero-mode F-spartners will acquire masses radiatively, primarily from gauge interactions.
The masses of these fields have been calculated in~{\cite{APQ}}.
\begin{eqnarray}
m_Q^2 &=& {\rm K} \; \frac{1}{ 4 \pi^4} \left( \frac{4}{3} g_3^2 + \frac{3}{4} g_2^2 + \frac{1}{36} 
g_1^2 \right) \frac{1}{R^2} \nonumber \\
m_U^2 &=& {\rm K} \; \frac{1}{ 4 \pi^4} \left( \frac{4}{3} g_3^2 + \frac{4}{9} g_1^2 \right) 
\frac{1}{R^2} \nonumber \\
m_D^2 &=& {\rm K} \; \frac{1}{ 4 \pi^4} \left( \frac{4}{3} g_3^2 + \frac{1}{9} g_1^2 \right)
\frac{1}{R^2} \nonumber \\
m_L^2 &=& {\rm K} \; \frac{1}{ 4 \pi^4} \left( \frac{3}{4} g_2^2 + \frac{1}{4} g_1^2 \right)
\frac{1}{R^2} \nonumber \\
m_E^2 &=& {\rm K} \; \frac{1}{ 4 \pi^4} g_1^2 \frac{1}{R^2}
\label{scalarmasses}
\end{eqnarray}
Here K is a dimensionless constant whose numerical value is close to 2.1, while $g_3$, $g_2$ and
$g_1$ are the gauge coupling constants of SU(3$)$, SU(2$)_{\rm L}$ and U(1$)_{\rm Y/2}$
respectively. Here we have neglected contributions to the masses of the F-spartners from their
Yukawa couplings to the Higgs, which are negligible except for the third generation F-stops. 
As shown in the appendix, for these fields we need to add
\begin{equation}
m_Q^2 = {\rm K} \; \frac{\lambda_t^2}{ 8\pi^4} \frac{1}{R^2}\, , \qquad
m_U^2 = {\rm K} \; \frac{ \lambda_t^2}{4  \pi^4}  
\frac{1}{R^2}\, .
\label{scalarmassesyukawa}
\end{equation}

The only F-partners with masses less than 50 GeV are the gluons of mirror color. These will confine into 
F-glueballs at a scale $\Lambda_{\rm F-QCD}$ of order a few GeV. Since these couple only very weakly to the 
SM particles at low energies, they evade current experimental bounds.
       
In this theory where do the leading contributions to the Higgs potential come from? As shown in the 
appendix gauge interactions give rise to a finite and positive one loop contribution to the mass parameters 
of both the up-type and down-type Higgs that takes the form
\begin{equation}
\label{gaugecontribution} 
\delta m_H^2|_{\rm gauge} = {\rm K} \; \frac{3 g_2^2 + g_1^2}{16 \pi^4} \frac{1}{R^2}
\end{equation}
However, at two loops
there is a finite negative contribution to the mass of the up-type Higgs from the top sector of 
order
\begin{equation}
\label{topcontribution}
\delta m_H^2|_{\rm top} \approx - \frac{3 \lambda_t^2}{ 4 \pi^2} \tilde{m}_t^2 \; {\rm log} \left( 
\frac{1}{R \; \tilde{m}_t} \right)
\end{equation}
where $\tilde{m}_t$ is the mass of the F-stop. A quick estimate suggests that 
the top contribution is larger in magnitude than the gauge
contribution from Eq.({\ref{gaugecontribution}}), leading to electroweak symmetry breaking.
However, to be certain of this a more careful analysis is required, which we leave for future
work.

At this stage the tree-level Higgs quartic in our model is identical to that of the MSSM, and is therefore 
too small to give rise to a Higgs mass larger than the current experimental lower bound of 114 GeV. We 
therefore extend the Higgs sector by adding to the theory an extra singlet $S$ which is localized to the 
brane at $y=0$ and couples to the Higgs as
\begin{equation}
\delta \left( y \right) \int d^2 \theta \left[\alpha S + \lambda S H_u H_d + \kappa 
S^3 \right]
\end{equation}
For ${\rm tan}\beta$ close to one and values of $\lambda$ greater than about 0.65 this will give rise to 
tree level Higgs masses greater than the experimental lower bound. Since the cutoff $\Lambda$ of the theory 
is low, it is not difficult to generate a value for $\lambda$ larger than this {\cite{TW}}. In the absence 
of the linear term in the superpotential the Higgs potential has an exact U(1$)_{\rm R}$ symmetry. The term 
$\alpha S$ breaks this continuous symmetry in the softest possible way leaving behind only a discrete $Z_4$ 
R symmetry. This suffices to ensure the absence of an unwanted Goldstone boson. We choose the value of 
$\alpha$ to be of order weak scale size to obtain consistent electroweak breaking. This choice is
technically natural. We leave the problem of naturally generating $\alpha$ of this size for future work.

The VEV of the singlet $S$ serves as an effective $\mu$ term. A negative mass for the scalar in $S$
can be generated by introducing into the bulk two SM singlet hypermultiplets $\hat{P}_A$ and
$\hat{P}_B$. The boundary conditions on these fields are such as to allow only a fermion zero mode
for each of $\hat{P}_A$ and $\hat{P}_B$. The bulk $Z_{AB}$ symmetry interchanges $\hat{P}_A$ and
$\hat{{P}}_B$. In addition, under the $Z_{AB}'$ symmetry, $\hat{P}_A$ and $\hat{{P}}_B$ are 
also interchanged.  
Then on the brane at $y=0$ we can write the interaction
\begin{equation}
\delta \left( y \right) \int d^2 \theta \left[ \lambda_P S P_A P_B + 
\mu_{P} \left({P_A}^2 + {P_B}^2 \right) \right]
\end{equation}
The effect of the coupling $\lambda_P$ is to generate a negative mass squared for the scalar in
$S$ at one loop. The theory is also invariant under a discrete symmetry, pedestrian parity
{\cite{pedestrian}}, under which $\hat{P}_A$ and $\hat{{P}}_B$ change sign but all other fields 
are
invariant. Pedestrian parity ensures that the zero-mode fermions in $\hat{P}_A$ and $\hat{P}_B$
are stable. These particles are potential dark matter candidates.

We are now in a position to understand the extent to which this model addresses the LEP paradox.
Since the largest contribution to the Higgs mass arises from the top sector, and assuming the
lightest neutral Higgs is SM-like and the other Higgs fields are significantly heavier, we can
estimate the fine-tuning by the formula
\begin{equation}
\frac{m_{H, {\rm phys}}^2}{2 \delta m_H^2|_{\rm top}} \times 100\%
\end{equation}
Here $m_{H, {\rm phys}}$ is the physical mass of the lightest neutral Higgs, and $\delta 
m_H^2|_{\rm top}$
is to be calculated from Eq.({\ref{topcontribution}}). For a compactification scale $1/R$ of order
5 TeV, a cutoff $\Lambda$ of order 20 TeV and $m_{H, {\rm phys}} = 115$ GeV the
fine-tuning is about $12\%$. The fine-tuning decreases for larger values of $m_{H, {\rm phys}}$ and
falls to about $40\%$ for $m_{H, {\rm phys}} = 200$ GeV. For comparison, the SM with a cutoff of
20 TeV is fine tuned at the  $0.1\%$ level for $m_{H, {\rm phys}} = 200$ GeV. A complete solution to the 
hierarchy problem may be obtained if there is a warped extra dimension~{\cite{RS}} in addition to the 
compact fifth dimension. Models of this type have been constructed in~{\cite{CN99}}.

In the absence of further interactions between the $A$ and $B$ sectors the lightest F-spartner, which is 
right-handed F-slepton, is absolutely stable and does not decay. In order to avoid the cosmological bounds 
on stable charged particles we add to the theory the non-renormalizable interactions
\begin{equation}
\label{Fdecay}
\delta \left( y \right) \int d^2 \theta \left( \frac{Q_A Q_A Q_A L_B}{\Lambda} + \frac{Q_B Q_B Q_B 
L_A}{\Lambda} \right)
\end{equation}
and
\begin{equation}
\label{moreFdecay}
\delta \left( y \right) \int d^2 \theta \left( \frac{U_A U_A D_A E_B}{\Lambda} + \frac{U_B U_B D_B
E_A}{\Lambda} \right)
\end{equation}
where we have suppressed the indices labeling the different generations. An F-slepton can then decay to 
three quarks and the LSP, which in this model is mostly Higgsino or singlino. F-baryons are also no longer 
stable, and decay before nucleosynthesis. The SM baryons, however, are still stable because decays to 
F-sleptons or F-leptons are kinematically forbidden. Although the interactions in Eq.~(\ref{Fdecay}) and 
Eq.~(\ref{moreFdecay}) give rise to flavor violating effects, these are small and consistent with current 
bounds. Precision electroweak constraints on this model are satisfied, as shown in the appendix. 

What are the characteristic collider signatures of this theory? The F-sleptons can be pair produced at the 
LHC through their couplings to the W,Z and photon. Each F-slepton decays to three quarks and the LSP. The 
high dimensionality of the relevant effective operator implies that for reasonable values of the couplings 
the F-slepton may travel anywhere from a few millimeters to tens of meters before decaying. The collider 
signatures are therefore expected to consist of either six jet events with displaced vertices, or highly 
ionizing tracks corresponding to massive stable charged particles.

For the F-squarks the situation is rather different. While they can also be pair produced in colliders 
through their couplings to the W,Z and photon they are charged under F-color rather than SM color. Then the 
absence of light states with charge under F-color other than F-gluons prevents the F-squarks from 
hadronizing individually. They therefore behave like scalar quirks~{\cite{Matt1, LKN}}, or `squirks'. The two 
F-squarks are connected by an F-QCD string and together form a bound state. This bound state system is 
initially in a very excited state but we expect that it will quickly cascade down to a lower energy state 
by the emission of soft F-glueballs and photons. Eventually the two F-squarks pair-annihilate into two (or 
more) hard F-glueballs, two hard W's, two hard Z's or two hard photons. They could also pair-annihilate 
through a single off-shell W,Z or photon into two hard leptons or jets. The decay of the bound state is 
prompt on collider time scales.

Before we can understand the collider signatures associated with the production of squirks, we must first 
estimate the F-glueball lifetime. Below the mass scale of the F-squarks this is essentially a `hidden 
valley' model~{\cite{Matt1, Matt2}}. The F-glueball must decay back to SM states because decays to F-(s)partners 
are kinematically forbidden. The dominant decays occur through an off-shell Higgs and the decay products 
are charm quarks and tau leptons, and perhaps bottom quarks as well if the F-glueball is sufficiently 
heavy. The coupling of the Higgs to the F-glueball is through a loop of virtual F-stops. The high 
dimensionality of this operator implies that F-glueballs are stable on collider time scales and almost all 
escape the detector. A naive estimate of the range yields about 10 Km., but this answer is very sensitive 
to the exact values of $\Lambda_{\rm F-QCD}$ and the F-glueball mass. As a consequence, it is conceivable 
that the range is as much as a factor of a thousand smaller than this, which would be very exciting from 
the collider viewpoint. Nevertheless, in what follows, we shall trust our naive estimate of the lifetime 
and assume that F-glueballs escape the detector.
  
The characteristic signatures of this scenario therefore include (but are not limited to) events with
\begin{itemize}
\item
four hard leptons (from the two Z's) accompanied by missing energy,
\item
two hard leptons (from the two W's, or from the off-shell Z or photon) accompanied by missing energy, 
and
\item
two hard photons accompanied by missing energy
\end{itemize}
It should be possible to determine the masses of the F-squarks from the energy distributions of the 
outgoing leptons. These signatures are very distinctive, and, if there are enough events, should make this 
model relatively straightforward to distinguish at the LHC. In addition to this, it may be possible to 
detect the characteristic experimental signatures of TeV size extra dimensions, such as Kaluza-Klein 
resonances and deviations from Newtonian gravity at sub-millimeter distances~{\cite{submm}}. The
cosmology of theories with a TeV size extra dimension has been considered, for example, in~{\cite{cosmo}}. 

Neutrino masses in this model may be either Dirac or Majorana. Since Majorana neutrino masses violate 
lepton number by two units, the $A \rightarrow B$ symmetry together with Eqns.~({\ref{Fdecay}}) and 
({\ref{moreFdecay}}) imply that in this case the model predicts neutron-antineutron oscillations. However, 
since the amplitude for this process is proportional to the small neutrino masses, the rates for this are 
completely consistent with the current experimental bounds~{\cite{nnbar}}.

\section{Conclusions}

In summary, we have constructed a new class of theories which address the LEP paradox. These
`folded supersymmetric' theories predict a rich spectrum of new particles at the TeV scale which
may be accessible to upcoming experiments. Together with mirror symmetric twin Higgs models, these
theories are explicit counterexamples to the conventional wisdom that canceling the one-loop
quadratic divergences to the Higgs mass parameter from the top sector necessarily requires new
particles charged under SM color.


\bigskip
\noindent {\bf Acknowledgments --} \\
We thank E. Cheu, M. Luty, T. Okui, M. Strassler and E. Varnes for discussions. G.B. acknowledges the 
support of the State of S\~{a}o Paulo Research Foundation (FAPESP), and of the Brazilian National 
Counsel for Technological and Scientific Development (CNPq). 
Z.C and H.S.G are supported by the NSF under grant 
PHY-0408954.  R.H. is supported by the US Department of Energy under Contract No. DE-AC02-76SF00515.

\bigskip

\appendix
\section{Radiative Corrections in Higher Dimensions}
\label{cancelation}

\subsection{The SU(3) $\times$ SU(3) Model}

In this appendix we will determine the radiative corrections to the Higgs mass in five dimensional 
folded-supersymmetric models. The corrections are finite and depend on the the supersymmetric structure of 
the higher dimensional theory as well as other global or gauge symmetries, depending on the specific model. 
We first focus on the realistic SU(3) $\times$ SU(3) model of section~\ref{realistic}. We shall see that 
the 
cancellation of one loop quadratic divergences is a consequence of supersymmetry and the discrete $Z_2$ 
symmetry $Z_{AB}$. We note that the calculation for the SU(6) model of section~\ref{top} is identical to 
the one outlined below.

The one loop effective potential for the Higgs has the convenient property that different interactions of 
the Higgs contribute additively. Therefore, when calculating the contribution to the potential from, say, 
the top Yukawa, one can ignore the gauge interaction of the Higgs and vice versa. 
Following~\cite{PeskinMirabelli, Jay5Dsusy} we write the part of the higher dimensional Lagrangian which is 
relevant for the cancellation of one loop quadratic divergences from the top sector in terms of 
$\mathcal{N}=1$ superfields. Even though the bulk Lagrangian possesses an SU(2$)_R$ symmetry, it is 
convenient to write the higher dimensional Lagrangian in the SU(2$)_R$ basis that is aligned with the 
unbroken $\mathcal{N}=1$ supersymmetry on the Higgs brane.
\begin{widetext}
\begin{eqnarray}
\mathcal{L}_{5D} &=& \int d^4 \theta \left[  Q_A^\dagger Q _A +
Q_A^{c\dagger} Q^c_A +U_A^\dagger U_A +U_A^{c\dagger} U^c_A  +
Q_B^\dagger Q _B + Q_B^{c\dagger} Q^c_B +U_B^\dagger U_B +U_B^{c\dagger} U^c_B 
\right] \nonumber  \\
&+& \int d^2 \theta \left [Q_A^c \partial_5 Q_A + U_A^c \partial_5 U_A  +
Q_B^c \partial_5 Q_B + U_B^c \partial_5 U_B
\right]  +\mbox{h.c.}  \nonumber \\ 
&+& \delta(y)\left\{  \int d^4\theta H_u^\dagger H_u +
\int d^2 \theta \left[  \lambda_t H_u Q_A U_A + \lambda_t H_u Q_B U_B\  \right] + \mbox{h.c} \right\}
\label{RestrictedL}
\end{eqnarray}
\end{widetext}
Here, and for the rest of this appendix, we have suppressed the label `3' denoting the third generation for simplicity. We have also neglected 
possible brane kinetic terms which we will come back to later.

It is straightforward to decompose the bulk fields into Kaluza-Klein modes according to their various 
boundary conditions. Zero modes exist only for the $(+,+)$ fields, the fermion components of $Q_A$ and 
$U_A$ as well as the scalar components of $Q_B$ and $U_B$. The relevant part of the Lagrangian for the zero 
mode fields alone, in components, is
\begin{eqnarray}
\mathcal{L}_{(0)} &=&
\mbox{kinetic terms} 
+ \lambda_t H_u q_{A0} u_{A0} + \mbox{h.c.}  \nonumber \\
&+& \lambda_t^2 |H_u|^2|\tilde q_{B0}|^2 +\lambda_t^2 |H_u|^2|\tilde u_{B0}|^2\,.
\end{eqnarray}
Notice that this part of the Lagrangian has an accidental supersymmetry. In particular, if we switch all 
labels, $A\leftrightarrow B$, on \emph{scalar fields only}, it appears to have an \emph{exactly} 
supersymmetric structure. Furthermore, the higher-dimensional supersymmetry together with the $Z_{AB}$ 
interchange guarantees that a regulator exists which preserves this accidentally 
supersymmetric structure of the zero-mode Lagrangian. (Metaphorically, the cutoffs of the fermion and 
scalar sectors are identical). This demonstrates that the extra-dimensional theory is indeed an ultraviolet 
completion for the 4D folded supersymmetric model. The one loop contribution of the zero mode top sector to 
the Higgs mass vanishes.

Now let us turn our attention to the part of the Lagrangian involving Kaluza-Klein modes which is relevant 
for the cancellation of one loop quadratic divergences from the top sector.
\begin{widetext}
\begin{eqnarray}
\mathcal{L}^{(KK)}
&=&  \sum_n    \left[   \mbox{kinetic terms} \right] \nonumber\\
&+&  \sum_n      \left[    \frac{n}{R} q_{An}^c q_{An}   
 + \frac{n}{R} u_{An}^c u_{An} 
+ \left(\frac{2n+1}{2R}\right)^2 |\tilde q_{An}|^2 
+ \left(\frac{2n+1}{2R}\right)^2 |\tilde q^c_{An}|^2 
+ \left(\frac{2n+1}{2R}\right)^2 |\tilde u_{An}|^2 
+ \left(\frac{2n+1}{2R}\right)^2 |\tilde u^c_{An}|^2   \right.  \nonumber\\
 &\   &\qquad  +  \left. \frac{2n+1}{2R} q_{Bn }^c q _{Bn}
+\frac{2n+1}{2R} u_{Bn }^c u _{Bn}
+ \left(\frac{n}{R}\right)^2 |\tilde q_{Bn}|^2 
+ \left(\frac{n}{R}\right)^2 |\tilde q^c_{Bn}|^2 
+ \left(\frac{n}{R}\right)^2 |\tilde u_{Bn}|^2 
+ \left(\frac{n}{R}\right)^2 |\tilde u^c_{Bn}|^2 \right]  \nonumber  \\
&+& \sum_{n,m} \left[ 
  \lambda_t H_u q_{Bn}u_{Bm}  
+\lambda_t^2 |H_u|^2 |\tilde q_{An}|^2
+\lambda_t^2 |H_u|^2 |\tilde u_{Am}|^2   
+\lambda_t \frac{2m+1}{2R} H_u \tilde q_{An}\tilde u^{c*}_{Am}
+\lambda_t \frac{2n+1}{2R} H_u \tilde q^{c*}_{An}\tilde u_{Am}
\right.   \nonumber\\
&\ & \ \ \ +\left. \lambda_t H_u q_{An} u_{Am}
+\lambda_t^2 |H_u|^2 |\tilde q_{Bn}|^2
+\lambda_t^2 |H_u|^2 |\tilde u_{Bm}|^2
+\lambda_t \frac{m}{R} H_u \tilde q_{Bn}\tilde u^{c*}_{Bm}
+\lambda_t \frac{n}{R} H_u \tilde q^{c*}_{Bn}\tilde u_{Bm} \right] \nonumber\\
&+& \sum_{n} \left[ 
%
%
\lambda_t H_u q_{An} u_{A0}
+\lambda_t^2 |H_u|^2 |\tilde q_{Bn}|^2
+\lambda_t^2 |H_u|^2 |\tilde u_{B0}|^2
+\lambda_t \frac{n}{R} H_u \tilde q^{c*}_{Bn}\tilde u_{B0} \right] \nonumber\\
&+& \sum_{m} \left[ 
%
%
\lambda_t H_u q_{A0} u_{Am}
+\lambda_t^2 |H_u|^2 |\tilde q_{B0}|^2
+\lambda_t^2 |H_u|^2 |\tilde u_{Bm}|^2
+\lambda_t \frac{m}{R} H_u \tilde q_{B0}\tilde u^{c*}_{Bm} 
\right]\,.
\label{KKL}
\end{eqnarray}
\end{widetext}
Here the sums over the Kaluza-Klein indices $n$ and $m$ begin at one\footnote{ Notice that the terms of 
the form $|H_u|^2|q_n|^2$ and $|H_u|^2|u_m|^2$ are summed over both $n$ and $m$, and thus appear in the 
Lagrangian with an infinite coefficient of $(\sum_n 1)$. In the higher dimensional Lagrangian these 
infinite coefficients arise as $\delta(0)$ when the brane auxiliary fields are solved for. 
In~\cite{PeskinMirabelli} (and also below) it was shown that these infinities are needed in order to get 
the appropriate cancellations in the supersymmetric limit.}. The cancellation of the one loop contribution 
to the Higgs mass will be apparent once we familiarize ourselves with equation~(\ref{KKL}). The first term 
in brackets following the kinetic terms contains mass terms for the Kaluza-Klein fields. The second term in 
brackets, with summation over both $n$ and $m$, is the set of interactions between the Higgs and 
Kaluza-Klein modes. The last two terms in brackets are interactions between the Higgs, a zero mode and a 
single Kaluza-Klein mode. The interactions have been grouped such that every line in~(\ref{KKL}) involves 
degenerate fermions and scalars. For example, the second term in brackets involves the fermions $q_{An}$ 
and $u_{Am}$ with masses $n/R$ and $m/R$ respectively. The same term involves the scalars $\tilde q_{Bn}$, 
$\tilde q^c_{Bn}$ with mass $n/R$ and the scalars $\tilde u_{Bm}$, $\tilde u_{Bm}^c$ with mass $m/R$. 
Furthermore, the structure of the interactions within each term in brackets is formally identical to that 
of a supersymmetric theory. Specifically, if as before we switch all labels, $A\leftrightarrow B$, on 
\emph{scalar fields only}, the relevant part of the Lagrangian appears to have an \emph{exactly} 
supersymmetric structure. Note however that this relabeling is not a symmetry of the full Lagrangian, once 
one takes into account the gauge interactions. The Higgs is ``fooled'' into living in an exactly 
supersymmetric theory but only at one loop. The diagrams responsible for the cancellation are shown in 
Figure~(\ref{fig-diagrams-top}).
\begin{figure} 
\begin{center} 
\includegraphics[width=0.45\columnwidth]{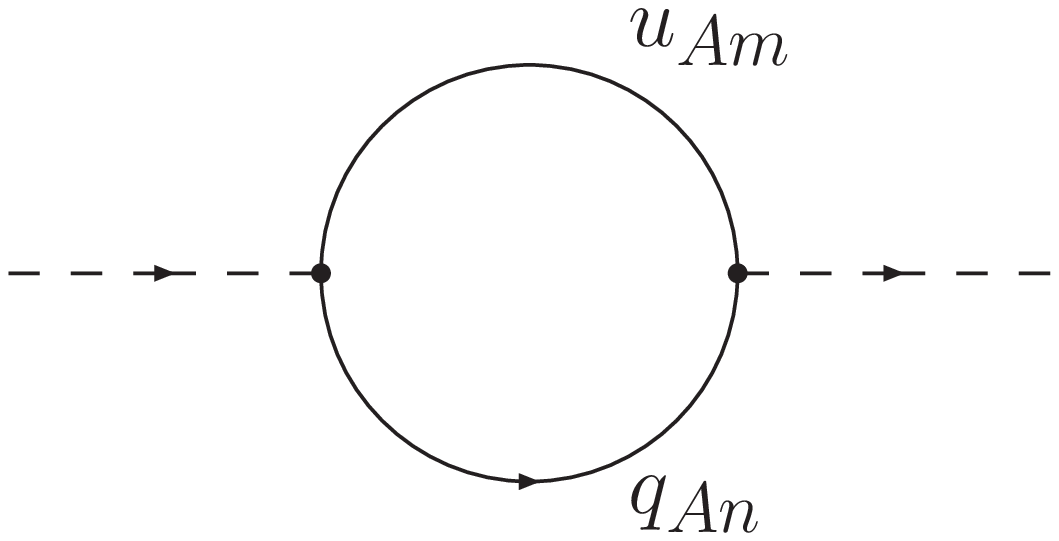}\\ 
\includegraphics[width=0.37\columnwidth]{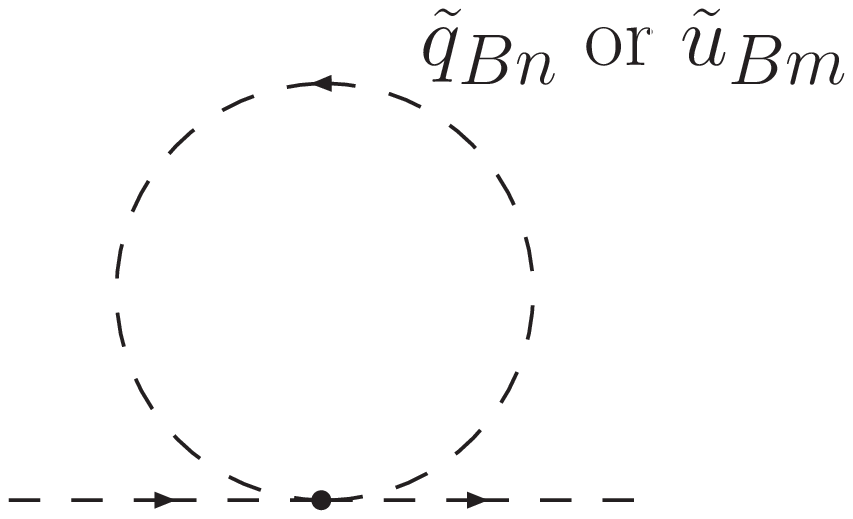} \ \ \ \ \ 
\ \ \ 
\includegraphics[width=0.45\columnwidth]{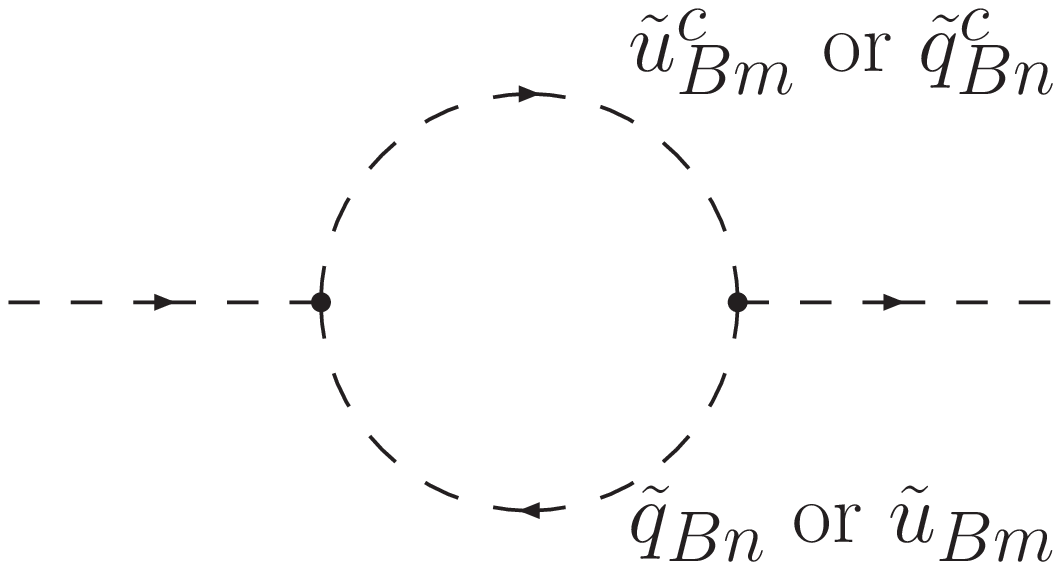} 
\end{center} 
\caption{The Feynman 
diagrams involved in the cancellation of the radiative corrections to the Higgs mass from the top sector. 
For every choice of two Kaluza-Klein levels $n$ and $m$ this combination of diagrams is present, adding up 
to zero. The cancellation has a supersymmetric form, but the scalar tops are not charged under SM color.} 
\label{fig-diagrams-top} 
\end{figure}

In addition to the terms in equation~(\ref{RestrictedL}) the Lagrangian may contain brane kinetic terms for 
the bulk fields. These break the bulk supersymmetry but preserve the appropriate 4D $\mathcal{N}=1$ 
supersymmetry on each brane. These terms can only be written for fields that are even around each fixed 
point, e.g. $Q$ at $y=0$ and $Q'$ at $y=\pi R$. However, by writing these brane kinetic terms explicitly 
one can verify that the accidental supersymmetry of the relevant part of the Lagrangian is preserved so 
long as the brane kinetic terms respect the unbroken $Z_2$ at each brane. We therefore require that the 
Lagrangians on the branes respect $Z_{AB}$ at $y=0$ and $Z_{AB}'$ at $y=\pi R$.
 
The exact cancellation of the one-loop Higgs mass in this model occurs only in the top sector. However, due 
to the non-local breaking of supersymmetry, the SU(2) gauge boson contribution is completely finite. To 
calculate 
this contribution we closely follow the approach of~{\cite{APQ}}. We are interested in the one loop 
Coleman-Weinberg (CW) effective potential~\cite{CW} due to gauge bosons, gauginos, as well as the chiral 
part of the $\mathcal{N}=2$ vector multiplet, running in loops. The CW potential is determined by 
calculating the one loop vacuum energy in terms of the field dependent masses. The CW potential from a 
Kaluza-Klein tower has the form
\begin{equation}
V(H)=\frac{1}{2}\mbox{Tr} \int \frac{d^4 k}{(2\pi)^4}
\sum_{n=-\infty}^{n=\infty} \ln \left(
\frac{k^2+ M^2_{Bn}+M^2(H)}{k^2+ M^2_{Fn}+M^2(H)}
\right)
\end{equation} 
where $M^2_{Bn}$ and $M^2_{Fn}$ are the $n$th boson and fermion Kaluza-Klein mass respectively, and 
$M(H)^2$ is the field dependent part of the mass which is common for both bosons and fermions due to 
supersymmetry.  Since we are only interested in the mass, we can focus on the coefficient of~$|H|^2$
\begin{eqnarray}
\label{CWmass}
m^2_H=&\mbox{Tr}& \frac{d M^2(H)}{d|H|^2} \\
&\times& \int \frac{d^4 k}{(2\pi)^4}
\sum_{n=-\infty}^{n=\infty} \left[ \frac{1}{k^2 + M^2_{Bn}}-\frac{1}{k^2 + M^2_{Fn}}\right] \nonumber
\end{eqnarray}
In our case, the fermions, whose boundary conditions are $(+,-)$ and $(-,+)$, have Kaluza-Klein masses of 
$(2n+1)/2R$. 
The bosons, with boundary conditions of $(+,+)$ and $(-,-)$ have Kaluza-Klein masses of $n/R$. However, the 
trace in 
equation~(\ref{CWmass}) runs only over $(+,+)$ and $(+,-)$ since only those couple to the Higgs brane.  
The sum over Kaluza-Klein modes is performed before 
integrating over phase space. 
One should note that the couplings of the Higgs to the Kaluza-Klein gauge bosons and gauginos is not 
diagonal in 
the Kaluza-Klein basis because the Higgs is localized to the brane. However, these off diagonal gauge 
interactions do not contribute to the Higgs mass at one loop, simplifying the calculation significantly.
The final result is for the one-loop contribution to the Higgs mass from SU(2) gauge bosons is
\begin{equation}
\label{massgauge}
\delta m^2_H|_{\mathrm{gauge}}= {7 C(H) \zeta(3)}\frac{g^2}{16 \pi^4 R^4}
\end{equation}
where $C(H)$ is the quadratic Casimir. Identifying $K\equiv 7 \zeta(3)/4\sim 2.1$ gives the one loop mass 
of equation~(\ref{gaugecontribution}). Finally, one can perform a similar calculation for the Yukawa 
contribution to the mirror stop masses (again neglecting off diagonal elements in the stop dependent mass 
matrix). This yields the stop mass contributions of equations~(\ref{scalarmasses}) 
and~(\ref{scalarmassesyukawa}).

\subsection{The SU(4) Model}

We now turn to the SU(4) model of section~\ref{gauge}. In this model the quadratic divergence that comes 
from loops of gauge boson zero modes is partially cancelled by loops of the zero modes of the off-diagonal 
gaugino bi-doublets. In the full extra dimensional theory the remaining quadratic divergence is cutoff at 
the scale $1/R$ because of the non-local nature of supersymmetry breaking in Scherk-Schwarz theories. In 
other words, when the sum over the entire Kaluza-Klein tower is added to the contribution of the zero mode 
to the Higgs mass, we get finite results. The extent of the cancellation between the boson Kaluza-Klein 
tower and the fermion Kaluza-Klein tower can be understood from the relevant group theory factors.  The 
group theory factor $\sum_a T^aT^a$ coming from the diagrams involving the diagonal block is +7/8 (which 
is 3/4 from the SU(2) plus 1/8 from the diagonal generator of SU(4)). The off-diagonal bi-doublets 
contribute a factor of $-1$ (the sign is opposite because the spin of the particles in this tower is 
opposite). The left over contribution thus has a coefficient of $-1/8$. Note that the residual factor of 
$1/8$ would be cancelled exactly by the additional diagonal gauge boson in the case where the full U(4) 
is gauged, as expected.

It is straightforward to explicitly sum the Kaluza-Klein towers using the methods of ~\cite{APQ}. 
The gauge contribution to the Higgs mass squared in this model is then
\begin{equation}
m^2_h|_{\rm gauge}= - {\rm K} \frac{g_2^2}{32\pi^4} \frac{1}{R^2} .
\end{equation}
This is about a factor of 20 smaller than the naive SU(2) gauge contribution 
to the Higgs mass squared in the SM,
\begin{equation}
m^2_h|_{\rm gauge, SM} \approx \frac{9 g_2^2}{64 \pi^2} \Lambda^2
\end{equation}
when cut-off at the scale $\Lambda = 1/R$. 

\section{Electroweak Precision Constraints}

Electroweak precision measurements tightly constrain the couplings of SM matter. The F-(s)partners and 
superpartners only contribute to these processes through loops, and therefore give only subdominant 
contributions. Thus for the most part, the theory we consider is the SM in a five dimensional bulk, with 
the Higgs localized at $y=0$.  In this theory Kaluza-Klein number is conserved in the 
bulk~\cite{uedbounds}. Violation of Kaluza-Klein number is induced by brane-localized operators, which are 
volume suppressed. As a consequence, the exchange of gauge Kaluza-Klein modes does not lead to new, 
unsuppressed, four-fermion operators of the zero-mode fermions. The volume suppression in these 
four-fermion operators renders their effects harmless.

Nevertheless, there remain several sources of deviations from the SM values. The main constraint on the 
compactification scale $R$ will come from the tree-level mixing between the gauge boson zero modes and the 
their Kaluza-Klein excitations. This is induced by the presence of the localized Higgs vacuum expectation 
value $v$, and results in tree-level contributions to the electroweak parameters $S$, $T$ and $U$.

In order to compute the electroweak constraint, we assume the limit
in which the light Higgs couples very nearly as the SM Higgs.
The Higgs is localized  on the brane at $y=0$.
We are interested in the gauge sector, including  the scalar kinetic term.
The relevant terms in the action are 
\begin{eqnarray}
S_{5D} &=& \int d^4x\int_0^{\pi R}  dy  \left\{ 
-\frac{1}{4}W_{MN}^aW^{aMN} -\frac{1}{4}B_{MN}B^{MN} \right.\nonumber\\ 
&&\left. 
+\delta(y) \left(D_\mu\Phi\right)^\dagger D^\mu\Phi
+ \cdots \right\}
\label{fds}
\end{eqnarray}
where $\Phi$ is the localized Higgs doublet and the covariant derivative is
\begin{equation}
D_\mu\Phi = \left(\partial_\mu -i g_5W_\mu^at^a 
-ig'_5\frac{1}{2}B_\mu\right)\Phi~.
\end{equation}
The expansion in Kaluza-Klein modes for $B_\mu(x,y)$ can be written as 
\begin{equation}
B_\mu(x,y) = \frac{1}{\sqrt{\pi R}}
\left\{B_\mu^{(0)}(x) + \sqrt{2}\sum_{n=1} B_\mu^{(n)}(x)
\cos\left(\frac{ny}{R}\right)\right\}~,
\label{kexp}
\end{equation}
and similarly for the $W_\mu^{a}$ fields. We work in the
$W_5^{a}=B_5=0$ gauge.
 
The vacuum expectation value of the Higgs   
\begin{equation}
\langle\Phi\rangle = 
\left(
\begin{array}{c}
0\\\frac{v}{\sqrt{2}}
\end{array}
\right)
\end{equation}
results in the 
localized action 
\begin{eqnarray}
S_{5D}^{\rm local.} &=& \int d^4x\int_0^{\pi R} dy \delta(y)
\frac{v^2}{4}\left\{g_5^2 W_\mu^-W^{\mu +} \right.\nonumber  \\
&& \left.+\frac{1}{2}(g_5^2+g_5^{'2}) Z_\mu
Z^\mu
\right\}
\end{eqnarray} 
leading to the masses of the $W$ and $Z$ zero modes. 
These terms also induce mixing between the zero modes and the 
Kaluza-Klein excitations. These are given by  
\begin{eqnarray}
W^{\pm(0)} \bullet  W^{\pm(n)} &=& \sqrt{2} \frac{g_2^2v^2}{4} \nonumber\\
Z^{(0)} \bullet Z^{(n)} &=& \sqrt{2} \frac{(g_2^2+g_1^{2}) v^2}{8}
\label{mixing}
\end{eqnarray}
where we have used $g_5=g_2\sqrt{\pi R}$ and $g'_5=g_1\sqrt{\pi R}$, defining
the 5D couplings. 
These mixings induce tree-level shifts in the $W$ and $Z$ wave-functions, 
resulting in contributions to the ``vacuum polarizations'' 
$\Pi_{WW}(q^2)$ and $\Pi_{ZZ}(q^2)$.  
In general, these can be expanded around low momenta:
\begin{equation}
\Pi_{VV'}(q^2) = \Pi_{VV'}(0) + q^2\,\Pi'_{VV'}(0) + \cdots
\end{equation}
We consider the following electroweak parameters~\cite{ewparam}: 
\begin{eqnarray}
\alpha\,S &=& 16\pi\,\left(\Pi'_{33}(0) - \Pi'_{3Q}(0)\right)\nonumber\\
&=& 4\,\sin^2\theta_W\, \cos^2\theta_W \, \Pi'_{ZZ}(0)~,
\label{sdef}
\end{eqnarray}

\begin{eqnarray}
\alpha\,T &=& 16\pi\,\left(\Pi_{11}(0) - \Pi_{33}(0)\right)\nonumber\\   
&=& \frac{\Pi_{WW}(0)}{M_W^2} - \frac{\Pi_{ZZ}(0)}{M_Z^2}~, 
\label{tdef}
\end{eqnarray}

\begin{eqnarray}
\alpha\,U &=& 16\pi\,\left(\Pi'_{11}(0) - \Pi'_{33}(0)\right)\nonumber\\   
&=& 4\,\sin^2\theta_W\,\left(\Pi'_{WW}(0) - \cos^2\theta_W\Pi'_{ZZ}(0)
\right)~.
\label{udef}
\end{eqnarray}
In the second lines of eqns.~(\ref{sdef}) and (\ref{udef}) we dropped 
terms proportional to $\Pi'_{\gamma\gamma}(0)$ and 
$\Pi'_{\gamma Z}(0)$ since there will be no contributions to them  
coming from the mixing in eqn.~(\ref{mixing}). Furthermore, when considering
loop contributions we can work in the $\overline{\rm MS}$ scheme, in which they are 
not present. 

In principle, we could include an extended electroweak parameter set, 
if we further expand the vacuum polarizations to order $q^4$. 
This results in four new parameters~\cite{Barbieri:2004qk}, 
involving the second derivatives
of the $\Pi_{VV'}(q^2)$. However, two of them correspond to 
combinations whose first derivatives were already considered leading to 
$S$ and $U$. In general, we expect the former to be suppressed with respect to 
the latter by  $M_Z^2/\Lambda^2$, with $\Lambda$ the scale of new physics.
This is specifically the case in our model.  The remaining two parameters
can be defined as 
\begin{eqnarray}
Y &=& \frac{M_W^2}{2} \, \Pi''_{BB}(0) 
= \frac{M_W^2}{2}\,\sin^2\theta_W\,\Pi''_{ZZ}(0)\label{ydef}\\
W&=& \frac{M_W^2}{2} \, \Pi''_{W_3W_3}(0)
= \frac{M_W^2}{2}\,\sin^2\theta_W\,\Pi''_{ZZ}(0)\label{wdef}~.
\end{eqnarray}
Once again, in the right-hand side of eqns.~(\ref{ydef}) and 
(\ref{wdef}) we ignored terms containing $\Pi''_{\gamma\gamma}(0)$ and 
 $\Pi''_{\gamma Z}(0)$. 

The main constraint from electroweak observables on the 
scale $R^{-1}$ comes from the mixing--induced contributions 
to the $T$ parameter. 
These give 
\begin{equation}
T \simeq -\pi \frac{(1-\tan^2\theta_W)}{\sin^2\theta_W}\,\zeta(2)\, 
(v\, R)^2 ~,
\label{t1}
\end{equation}
where we used the approximation $R^{-1} \gg M_{W,Z}$ and summed over an 
infinite number of Kaluza-Klein modes. 
This results in 
\begin{equation}
T \simeq -\frac{\pi^3}{6}\,\frac{(1-\tan^2\theta_W)}{\sin^2\theta_W}
\,(v\, R)^2 ~,
\label{t2}
\end{equation} 
giving 
\begin{equation}
T \simeq - 16 \,(v\, R)^2 ~.
\label{t3}
\end{equation} 
For instance, for $R^{-1}=5~$TeV, this gives $T \simeq - 0.04$, well 
in agreement with data. The current PDG best fit with $T$ and $S$ free gives
$T= -0.17\pm 0.12$ at $90\%$ C.L. However, given that this model gives a very
small contribution to $S$ (see below), the $90\%$ C.L. lower bound on
$T$ is about $-0.15$. This translates approximately into 
$R^{-1} >2.5~$TeV as the lower bound. 
\vskip0.5cm
The tree level mixing leads to contributions to the $S$ parameter  given
by 
\begin{eqnarray}
S &\simeq & - \frac{4\,\sin^2\theta_W\,\cos^2\theta_W}{\alpha}\,
\zeta(4)\, \left(M_Z\,R\right)^4 \nonumber\\
&\simeq & -100\,\left(M_Z\,R\right)^4~.
\label{scont}
\end{eqnarray}
This gives a very small value of $S$ for any sensible value of $R^{-1}$.

Finally, the contributions to the $U$ parameter give
\begin{eqnarray}
U &\simeq & -4\sin^2\theta_W\,cos^2\theta_W\,(\cos^2\theta_W - \sin^2\theta_W)
\nonumber\\
&&\times\,\zeta(4)\,(M_Z R)^4~,
\label{ucont}
\end{eqnarray}
which is of the same order as the $S$ contributions, and equally negligible.

There are also one-loop contributions 
of the Kaluza-Klein spectrum to electroweak parameters. These will be small since
the Kaluza-Klein states decouple in the $R^{-1}\gg v$ limit. 
The $T$ parameter gives the largest such contribution coming from 
the propagation through the Kaluza-Klein spectrum 
of the isospin breaking between the top and bottom 
Yukawa couplings. If we sum over all Kaluza-Klein modes the 
contribution to $T$ is approximately given by~\cite{uedbounds} 
\begin{equation}
T \simeq \frac{m_t^2}{8\pi^2 v^2\alpha}\, 
(m_t R)^2\,\zeta(2)
\simeq O(1) (v R)^2~,
\label{topoloop}
\end{equation}
which is about one order of magnitude smaller than the tree-level
contribution of eqn.~(\ref{t3}).  
The contributions from gauge Kaluza-Klein modes are considerably more suppressed. 

There are also loop contributions to the $S$ parameter. However, these
will not result in a constraint on $R^{-1}$ once the bounds are satisfied
by the loop contributions to $T$. The reason for this is that since
Kaluza-Klein fermions are vector-like, they can only contribute to $S$ through
large mass splitting, which would result in an even larger contribution
to $T$. For instance, summing the leading contribution coming from the 
top-quark Kaluza-Klein modes, one gets
\begin{equation}
S \simeq 0.01\, \zeta(2)\, (m_t R)^2~,
\label{s1}
\end{equation}
which is negligibly small for any realistic value of $R^{-1}$. 

Contributions from Kaluza-Klein loops to the $U$ parameter are also negligible, since 
they require isospin violation of a size forbidden  by the $T$ parameter
constraint.

Finally, we consider the effects of the mixing induced between 
zero-mode fermions and their Kaluza-Klein excitations, after electroweak symmetry 
breaking. This comes from the localized Yukawa couplings
\begin{equation}
\int d^4 x\int dy \delta(y)\,
(\pi R)\,\bar Q_L(x,y) H(x) Y_q q_R(x,y)~,
\label{yukawa}
\end{equation}
where the dimensionless matrix $Y_q$ is the four-dimensional one, and 
there will be a similar term for leptons. 
The resulting mixing goes like
\begin{equation}
q^{(0)}\bullet q^{(n)}  =  \sqrt{2} \, m_q~,
\label{fermimix}
\end{equation}
and similarly for leptons. The couplings of gauge bosons with fermions will
then suffer flavor dependent shifts due to this mixing. 
The largest effect will be in the $Z$ coupling with the top quark. 
However, this will not be accurately known  
any time soon.
On the other hand, the couplings to the $b$ quarks are very well measured.
For instance, the induced shift in the coupling of the 
$Z$ to left-handed $b$ quarks, the best known of the $b$ couplings, is 
approximately 
\begin{equation}
 \delta g_L^b \simeq 2\,\sqrt{2}\,\zeta(2)\,(m_b R)^2\,g_R^b ~,
\label{deltabl}
\end{equation}
where once again we summed over all the Kaluza-Klein modes. 
For instance, if $R^{-1}=1~$TeV, the effect is $O(10^{-5})$, well 
bellow the deviation allowed by experiment, which must satisfy
 $\delta g_L^b/g_L^b < 0.01$ if we consider a $3\sigma$ interval whithin the 
experimental determination at LEP. 
Furthermore, the shifts of the couplings to lighter quarks will 
be un-observably small. 

The mixing in eqn.~(\ref{fermimix}), and its
flavor-dependent effects in the couplings to the $Z$, lead in principle
to flavor violation in the $Z$ interactions, and therefore 
to flavor changing neutral currents (FCNC) mediated by the $Z$. 
These will be suppressed by a factor of $(m_q R)^4$. 
As an illustration, we show the contribution to $B_d^0-\bar{B_d^0}$ mixing. 
This is given by 
\begin{equation}
{\cal H}_{\rm f.v.}^{\Delta B=2} \simeq 0(1)\,(D_L^{bd})^2\,
\frac{8}{\sqrt{2}}\,G_F\,(m_b R)^4\,Q~,
\label{fvkkmix}
\end{equation}
where  
$Q=\bar{b}(1-\gamma_5)d\,\bar{b}(1-\gamma_5)d$ is the operator responsible 
for the $\Delta B =2$ transition, and $D_L^{bd}$ is an element of the 
matrix rotating left-handed 
down quarks from the weak to the mass eigen-basis. Naively, we expect 
this to be $O(\sin^3\theta_c)$, with $\theta_c$ the Cabibbo angle.
We compare this with the CP conserving 
SM contribution coming from the box diagrams, mainly
containing the top quark. This is approximately given by  
\begin{equation}
H_{\rm SM}^{\Delta B=2} = \frac{G_F^2}{16\pi^2}\,M_W^2\,(V^*_{tb}V_{ts})^2\,
\eta_{B} S_0(x_t)\,Q~,
\label{smkmix} 
\end{equation}
where $S_0(x_t)$ is the loop function depending on $x_t = (m_t/M_W)^2$, 
$\eta_B$ is a QCD correction and $\eta_B S_0(x_t)\simeq 1$. 
We see that the coefficient of the flavor violating contribution 
will be about two orders of magnitude smaller  than the SM one, 
for any value of $R^{-1}$ allowed by other 
electroweak precision constraints, mainly $T$. This will give
\begin{equation}
{\cal H}_{\rm f.v.}^{\Delta B=2}\simeq 
2.6\times 10^{-16} {\rm GeV}^{-2}\,(D_L^{bd})^2
\,\left(\frac{2~{\rm TeV}}{R^{-1}}\right)^4\,Q~,
\label{fvkkmix}
\end{equation}  
which is at least two orders of magnitude below the SM value, even
if we allow for $D_L^{bd}\simeq O(1)$. If however, we consider the 
standard ansatz $D_L \simeq V_{\rm CKM}$, $D_L^{bd} \simeq V_{bc}$, 
the flavor violating contribution drops another three orders of magnitude.
For the case of $B_s^0-\bar{B_s^0}$ mixing, the effect will be 
four orders of magnitude smaller than the SM, assuming $D_L^{bs}\simeq O(1)$.
In $K^0-\bar{K^0}$ mixing the strange quark mass suppresses the effect even 
further. 

Finally, the non-universal contribution to the $Z t\bar t$ coupling
results in a $t c Z$ vertex, leading to a tree-level contribution 
to the rare top decay $t\to c Z$ giving 
\begin{eqnarray}
BR(t\to c Z)&\simeq &(U^{tc})^2\, \left(m_t R\right)^4 \nonumber \\
&\simeq & (U^{tc})^2\, 6\times 10^{-5} 
\left(\frac{2~{\rm TeV}}{R^{-1}}
\right)~,
\label{brtcz}
\end{eqnarray}
where we defined $(U^{tc})^2 = (U_L^{tc} \,g_{Lt})^2 + (U_R^{tc} \,g_{Rt})^2$, 
in terms of the left and right-handed
rotation matrix elements. Thus, if the matrix elements of the rotation 
of up quaks to the mass eigen-basis is not very small, these 
branching ratios could be observable at the LHC, where a sensitivity of $10^{-5}$ in these decay modes can be 
achieved for this low value of $R^{-1}$. However, more generically, these rare process is highly suppressed.

\end{document}